\documentclass{agujournal2019}
\usepackage{url} 
\usepackage{lineno}

\usepackage[authormarkup=none,final]{changes}
\definechangesauthor[name=editor, color=red]{r0}
\definechangesauthor[name=editor, color=orange]{r1}
\definechangesauthor[name=editor, color=cyan]{r2}
\definechangesauthor[name=editor, color=purple]{r3}
\newcommand{\revise}[1]{{\color{black}{#1}}}

\newcommand{\bx}[0]{\mathbf{x}}
\newcommand{\RR}[0]{\mathbf{R}}

\usepackage{soul}
\linenumbers
\usepackage{amsmath,bm}
\usepackage{lmodern}
\usepackage{amsthm}

\newtheorem*{remark}{Remark}
\newtheorem*{example}{Example}

\usepackage{todonotes} 
\usepackage[caption=false]{subfig}
\journalname{Water Resources Research}

\usepackage{bm}
\usepackage{cleveref}

\usepackage{booktabs}
\nolinenumbers
\begin{document}

%
%


\title{\replaced{Coupled Time-lapse Full Waveform Inversion for Subsurface Flow Problems using Intrusive Automatic Differentiation}{Time-lapse Full Waveform Inversion for Subsurface Flow Problems with Intelligent Automatic Differentiation}}

%
%




\authors{Dongzhuo~Li\affil{1}\(^\ast\), Kailai~Xu\affil{2}\thanks{Both authors contributed equally to this work.}, Jerry~M.~Harris\affil{1,2}, Eric~Darve\affil{2,3}}


\affiliation{1}{Department of Geophysics, Stanford University, Stanford, CA, 94305}
\affiliation{2}{Institute for Computational and Mathematical Engineering, Stanford University, Stanford, CA, 94305}
\affiliation{3}{Mechanical Engineering, Stanford University, Stanford, CA, 94305}




\correspondingauthor{Dongzhuo Li}{lidongzh@stanford.edu}




\begin{keypoints}
\item We assimilated seismic waveform data to directly invert for \replaced{hydrological subsurface properties (e.g., permeability)}{intrinsic parameters of subsurface flow (e.g., permeability)}.
\item Coupled inversion leads to more accurate inversion result compared to decoupled inversion. 
\item We adopted an \replaced{intrusive}{intelligent} automatic differentiation strategy with \replaced{custom}{customized} operators for high computational efficiency and scalability.
\end{keypoints}

%
%


\begin{abstract}
We describe a novel framework for \replaced{estimating subsurface properties, such as rock permeability and porosity, from time-lapse observed seismic data by coupling full-waveform inversion, subsurface flow processes, and rock physics models.}{PDE (partial-differential-equation)-constrained full-waveform inversion (FWI) that estimates parameters of subsurface flow processes, such as rock permeability and porosity, using time-lapse observed data}. \deleted{The forward modeling couples flow physics, rock physics, and wave physics models.} For the inverse modeling, we handle the back-propagation of gradients by an \replaced{intrusive automatic differentiation}{intelligent automatic differentiation} strategy that offers three levels of user control: (1) \replaced{a}{A}t the wave physics level, we adopted the discrete adjoint method in order to use our existing high-performance FWI code; (2) at the rock physics level, we used built-in operators from the \texttt{TensorFlow} backend; (3) at the flow physics level, we \replaced{implemented}{code} customized PDE operators for the potential and nonlinear saturation equations\deleted{, which highly resemble neural network layers}. These three levels of gradient computation strike a good balance between computational efficiency and programming efficiency, and when chained together, constitute a coupled inverse system. We use numerical experiments to demonstrate that (1) the three-level coupled inverse problem is superior in terms of accuracy to a traditional \added{decoupled inversion} strategy; (2) it is able to simultaneously invert for parameters in empirical relationships such as the rock physics models; and (3) the inverted model can be used for reservoir performance prediction and reservoir management/optimization purposes.
\end{abstract}



%
%

%


%
%
%
%

\section{Introduction}
Accurate inversion for the parameters that govern subsurface flow dynamics plays a vital role in addressing energy and environmental problems, such as oil and gas recovery~\cite{teletzke2010enhanced}, $\text{CO}_2$ storage~\cite{shi2013snohvit}, salt-water intrusion~\cite{beaujean2014calibration}, and groundwater contaminant transport~\cite{reid1996functional,mclaughlin1996reassessment}. A calibrated flow model can help optimize hydrocarbon production, evaluate environmental risks, and remediate environmental damage.

\deleted{There are three types of material properties in this problem: (1) intrinsic rock properties, such as permeability and porosity, related to fluid flow and storage; (2) flow properties such as saturation and pressure; (3) elastic properties such as bulk and shear moduli. Those properties are connected by models of the flow physics, rock physics, and wave physics as illustrated in Fig.~1. In the aforementioned applications, the inverse problem is to use observed data (e.g., seismic) to estimate intrinsic rock properties, subject to constraints from the physics models. Typically, the observed data can be classified into two types:}


\deleted{There are inversion challenges with both of the two types of data. On the one hand, traditional methods like history matching use direct measurements or in-situ flow data (e.g., production history and flow pressure in wells)} \revise{Conventional direct inversion methods (e.g., history matching~\cite{chen1974new,carter1974performance,chavent1975history}) rely on in-situ flow data, such as production history and flow pressure in wells,} to fit a flow equation for intrinsic properties such as permeability. However, flow data can only be collected from sparsely distributed wells, and the data are usually integrated along the wellbore. Thus, the inversion suffers from large null-space, and the results exhibit reduced resolution\revise{~\cite{yeh1986review}}.

\deleted{On the other hand, indirect measurements or soft data such as seismic waves, provide a high volumetric sampling of the reservoir and can reconstruct high-resolution models of elastic properties. However, they do not directly detect the intrinsic properties of interest. Hence, most conventional uses of time-lapse seismic data will first invert for elastic properties that are qualitatively interpreted for the intrinsic rock properties.}

\added{On the other hand, indirect geophysical measurements such as seismic waveform data, provide better volumetric sampling of the reservoir and can reconstruct high-resolution models of elastic properties.} The seismic data\deleted{~(indirect measurement)} are collected through time-lapse surveys, e.g., regular or irregular recordings made months or years apart. To be more specific, fast-time refers to the digital sampling interval of the repeated wavefield recordings, whereas slow-time refers to the evolution time or history of flow in a reservoir. The terms, ``fast-time'' and ``slow-time'',  are used to emphasize the different time scales for seismic surveys (e.g., days or weeks) and subsurface flow processes (e.g., months or years). In the slow-time case, the flow properties are assumed not to change, or change negligibly, in the short period during which seismic snapshots are recorded. We also assume that intrinsic properties such as permeability and porosity do not change with slow-time.

\revise{The main challenge of using seismic data is that they do not directly detect the intrinsic properties of interest. Hence, conventional uses of (time-lapse) seismic data first invert for elastic properties and then treat those intermediate properties as data to estimate intrinsic rock properties. We refer to this strategy as the decoupled inversion, which was for example proposed by \citeA{rubin1992mapping} and \citeA{copty1993geophysical}, and was also formulated as the seismic-history matching method~\cite{landa1997procedure,huang1998improving,emerick2007history,volkov2018gradient}. Although this approach improves the well-posedness of the problem and increases resolution, inaccuracies from the intermediate step of seismic inversion yield artifacts in inverted intrinsic properties~\cite{day2004assessing}.} 


\revise{Besides, the seismic inversion techniques used in subsurface flow problems are usually based on high-frequency approximation (ray theory) and simple linear convolution models. Those methods have the merit of being simple and robust, but they lead to reduced resolution because of limited information used. They also suffer from errors caused by inaccurate physics. Recent developments in FWI have provided great improvements in estimates of elastic properties with high spatial resolution and quantitative accuracy~\cite{tarantola1984inversion,virieux2009overview}. In hydrogeophysical applications, \citeA{klotzsche2012crosshole}, \citeA{klotzsche2014detection}, \citeA{gueting2015imaging,gueting2017high}, and \citeA{keskinen2017full} also find that ground-penetrating radar FWI yields a higher resolution result than ray-based GPR traveltime inversion. Thus, a coupled seismic inversion method fully based on wave equations in line with FWI is highly desirable.}

We \replaced{propose}{describe} herein \deleted{an}\added{a coupled} inversion framework that assimilates seismic waveform data to\deleted{directly and} \deleted{automatically} invert for the intrinsic parameters of subsurface flow \added{automatically}. This inversion approach combines flow physics, rock physics, and wave physics models. We consider a concrete example with the following three key physical components: 
 \begin{enumerate}
     \item  slow-time subsurface flow model $\mathcal{S}$: an immiscible incompressible two-phase flow model that maps permeability to evolution of fluid saturation and pressure;

     \item rock physics model $\mathcal{R}$: a patchy-saturation model or a Gassmann's model that maps saturation to rock elastic properties;
     
     \item fast-time seismic wave physics model $\mathcal{F}$: elastic wave propagation that maps elastic properties to seismic waveforms recorded at receiver arrays.
  \end{enumerate}
\deleted{(1) slow-time subsurface flow model $\mathcal{S}$: an immiscible incompressible two-phase flow model that maps permeability to evolution of fluid saturation and pressure; (2) rock physics model $\mathcal{R}$: a patchy-saturation model or a Gassmann's model that maps saturation to rock elastic properties; and (3) fast-time seismic wave physics model $\mathcal{F}$: elastic wave propagation that maps elastic properties to seismic waveforms recorded at receiver arrays.} \revise{In the field of hydrogeophysics, there has also been substantial research on the coupled inversion~\cite{linde2016joint}. In near-surface surveys, the geophysical data are usually electrical resistance tomography (ERT) data or ground penetrating radar (GPR) traveltime. The subsurface flow processes are often tracer dynamics in steady flow~\cite{irving2010stochastic,pollock2010fully,jardani2013stochastic} and unsaturated flow in the vadoes zone~\cite{kowalsky2004estimating,kowalsky2005estimation,hinnell2010improved,huisman2010hydraulic}. Petrophysical models such as Archie's law~\cite{archie1942electrical} interlink the hydrological flow properties and geophysical properties. In the above examples, the geophysical forward modeling is not transient by nature or simplified as so. For example, the ERT modeling is governed by the Poisson equation, and the GPR traveltime is computed by (straight) ray tracing~\cite{kowalsky2004estimating,kowalsky2005estimation}. Sometimes the transient flow equation is also transformed into a steady-state equation~\cite{pollock2010fully,pollock2012fully} to reduce the computation complexities. In our formulation, however, both the subsurface flow and the wave propagation are governed by time-dependent PDEs. The complex coupling makes it challenging to use the adjoint-state method~\cite{sun1990coupled}. Since our problem is inherently large-scale, it is also infeasible to use sampling-based methods or global optimization methods as many of the above examples did.}

\replaced{To meet this computational challenge}{Next}, we \replaced{propose}{add} \added{an} \replaced{intrusive}{intelligent} automatic differentiation \added{method to solve this coupled inversion problem} \replaced{with}{for} high computational efficiency and scalability. \replaced{We formulate the inverse problem as}{Our goal is to solve} a large-scale partial-differential-equation (PDE) constrained \replaced{optimization}{inverse} problem \replaced{to be solved by}{with} gradient-based optimization methods. However, deriving and implementing the gradients for this coupled system are both nontrivial and laborious. To overcome this difficulty, we \replaced{develop}{propose} an automatic\added{-}differentiation\replaced{-based method, intrusive automatic differentiation (IAD), which allows for computing gradients algorithmically and accurately}{ strategy with three levels of complexity}. \added{Our automatic-differentiation-based method is intrusive in the sense that we incorporate customized gradient back-propagation implementations, such as adjoint state or checkpointing schemes codes, into an automatic differentiation framework.  This technique allows for gradient back-propagation of custom-designed PDE solvers, even for implicit solvers, which typical automatic differentiation tools cannot handle. Another advantage of IAD is that the custom operators, such as a GPU-accelerated FWI module, enables high performance large-scale scientific computing. This set of tools will potentially benefit researchers in the hydrological community, as well as the inverse modeling community in a broader context.} \deleted{For flow physics models, we use customized PDE operators for one single step in time; for rock physics models, we use built-in operators in an automatic differentiation framework (\texttt{TensorFlow}); for the wave physics models, we adopt the discrete adjoint method in order to incorporate an efficient GPU-accelerated FWI code.}

In numerical examples, we follow a setup in $\text{CO}_2$ sequestration, in which supercritical $\text{CO}_2$ is injected into a reservoir filled with water. We \replaced[id=r2]{show in those examples that}{argue and present evidence that}
 \begin{enumerate}
     \item coupled inversion is more accurate than the traditional decoupled inversion, and

     \item the inversion can simultaneously invert for empirical closure relationship (e.g., the rock physics model parameters) and intrinsic parameters.
  \end{enumerate}
 
The paper is organized as follows. We first describe the general mathematical formulation of the PDE-constrained inversion problem and coupled physics models. Next, we present the \replaced{intrusive}{intelligent} automatic differentiation method. Finally, we conduct numerical examples, discuss, and offer conclusions.

\section{Problem Setup}\label{sect:problem_setup}
In this section, we first give a general formulation of the \added[id=r0]{coupled}\deleted[id=r0]{ time-lapse seismic} full-waveform inversion \replaced[id=r0]{method}{problem} for parameters in dynamic geophysical processes. We then discuss in detail the governing equation for the wave propagation and two-phase flow in porous media along with the rock physics that couple flow physics to wave physics. 

\subsection{General Formulation}\label{sect:general}

\replaced[id=r1]{The problm is}{Mathematically speaking, the problem can be} formulated as a PDE-constrained inverse problem. There are two physical processes governed by different PDEs. The first process, the slow-time dynamic flow process, is described by the equation
\begin{equation}\label{equ:slow}
    \frac{\partial \mathbf{c}(\mathbf{x}, t)}{\partial t} = \mathcal{S}\big(\mathbf{c}(\mathbf{x},t), K(\mathbf{x})\big),
\end{equation}
where $\mathbf{c}$ is the flow property, i.e., fluid saturation, \(K\) is the intrinsic rock property, i.e., permeability. Fluid saturation $\mathbf{c}$ will affect the elastic properties \(\mathbf{m}\), e.g., rock bulk modulus and density through a closure relationship  from the rock physics model $\mathcal{R}$:
\begin{equation}
    \mathbf{m}(\mathbf{x},t) = \mathcal{R}\big(\mathbf{c}(\mathbf{x},t)\big).
\end{equation}
The slow-time flow process is monitored indirectly within measurements of duration \([t_i, t_i + \Delta T_f]\), where \(t_i\) is the starting slow-time of the \(i\)-th measurement and \(\Delta T_f\) is the fast time interval. Each measurement is a seismic survey conducted by sequentially exciting sources and recording seismic data $\mathbf{d}^{\mathrm{obs}}_i$ at corresponding receiver positions. We can also compute predictions of wavefield by solving the wave equation \added[id=r2]{in the first-order velocity-stress form}
\begin{equation}\label{equ:fast}
      \frac{\partial \mathbf{u}_i(\mathbf{x}, t)}{\partial t} = \mathcal{F}\big(\mathbf{u}_i(\mathbf{x}, t), \mathbf{m}(\mathbf{x},t)\big),
\end{equation}
with \(\mathbf{m}\) as the assumed parameter. Here $\mathbf{u}$ is the wavefield, which for example can be the field of particle velocity or stress. The time duration of the process Eq.~(\ref{equ:fast}) is short compared to Eq.~(\ref{equ:slow}), i.e., $\Delta T_f\ll T_s$ and therefore we assume that $\mathbf{m}$ is fixed during \([t_i, t_i + \Delta T_f]\) and denote $\mathbf{m}_{t_i}(\mathbf{x}) = \mathbf{m}(\mathbf{x}, t_i)$. The fast-time scale equation reduces to 
  \begin{equation}
      \frac{\partial \mathbf{u}_i(\mathbf{x}, t)}{\partial t} = \mathcal{F}\big(\mathbf{u}_i(\mathbf{x}, t), \mathbf{m}_{t_i}(\mathbf{x})\big)
  \end{equation}
In the same manner of recording observed data, we sample the synthetic wavefield with operator \(\mathcal{Q}\): $\mathbf{d}_{i}=\mathcal{Q}\mathbf{u}_i$. We can then compute the misfit of synthetic data $\mathbf{d}$ and observed data ${\mathbf{d}}^{\mathrm{obs}}_i$ as \(\mathcal{J}(\mathbf{d}) = \frac{1}{2} \|\mathbf{d}_i^{\text{obs}} - \mathbf{d}_i\|^2\).

\begin{figure}[hbt]
\centering
  \includegraphics[width=\textwidth]{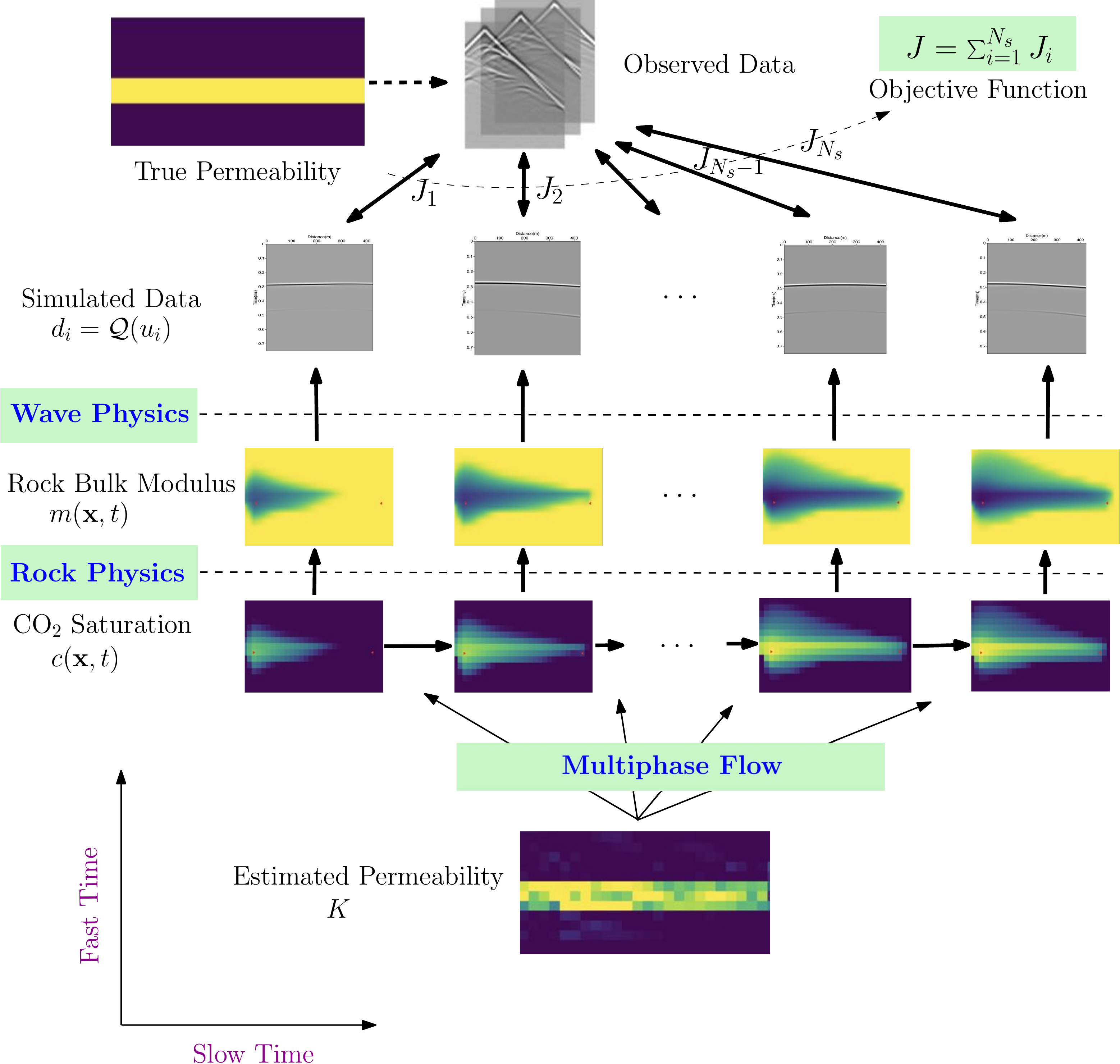}
  \caption{Schematic illustration of the coupled inverse problem.}
  \label{fig:backward}
\end{figure}

Putting together the above equations with proper boundary and initial conditions $\mathcal{B}_{\mathcal{F}}$ and $\mathcal{B}_{\mathcal{S}}$, we have
\begin{align}
     \min_{K} \ & J = \sum_{i=1}^{N_s} J_i = \frac{1}{2}\sum_{i=1}^{N_s} \|\mathbf{d}^{\text{obs}}_i-\mathbf{d}_i\|^2\\
     \mathrm{s.t.} \ & \mathbf{d}_i(\mathbf{x}, t) = \mathcal{Q}\big(\mathbf{u}_i(\mathbf{x}, t)\big)\label{equ:equ:J} \\
     & \frac{\partial \mathbf{u}_i(\mathbf{x}, t)}{\partial t} = \mathcal{F}\big(\mathbf{u}_i(\mathbf{x}, t), \mathbf{m}_{t_i}(\mathbf{x})\big)\quad \mathcal{B}_{\mathcal{F}}(\mathbf{u}_i) = 0 & t\in [t_i, t_i+\Delta T_{f}]\\
     & \mathbf{m}(\mathbf{x},t) = \mathcal{R}\big(\mathbf{c}(\mathbf{x},t)\big)\\
     & \frac{\partial \mathbf{c}(\mathbf{x}, t)}{\partial t} = \mathcal{S}\big(\mathbf{c}(\mathbf{x},t), K(\mathbf{x})\big)\quad \mathcal{B}_{\mathcal{S}}(\mathbf{c}) = 0 & t\in [0, T_s].
\end{align}
The coupled inversion is schematically illustrated in Fig.~\ref{fig:backward}.

\subsection{{Multi-phase Flow in Porous Media}}\label{sect:multi}
The physics of multi-phase flow in porous media is fundamental to many geoscience fields, such as hydrology, reservoir engineering, glaciology, and volcanology. It \replaced{underlies}{underlines} the slow-time-scale processes of geophysical time-lapse monitoring problems. As an example, we use the physical model of two-phase immiscible incompressible flow in this paper. This model can describe, at least to the first order, the phenomenon of the displacement of one fluid of the other. Examples include injection of water in a reservoir to produce oil, and injection of supercritical $\text{CO}_2$ into saline aquifers. \added{All symbols in the this model are listed in \Cref{tab:flow_physics_symbols}.}

\begin{table}[htpb]
\caption{Notation for the two-phase flow model. \added[id=r2]{The symbols $\mathbf{x}$ and $t$ denote space and time variables, respectively.}}
\centering
\begin{tabular}{l r}
\hline
 Symbol & Meaning  \\
\hline
  $\phi\added[id=r2]{(\mathbf{x})}$  & porosity   \\
  $K\added[id=r2]{(\mathbf{x})}$			& permeability \\
  $k_{ri}$ & relative permeability of fluid $i$ \\
  $S_i\added[id=r2]{(\mathbf{x}, t)}$  & saturation of fluid $i$   \\
  $P_i\added[id=r2]{(\mathbf{x}, t)}$			 & pressure of fluid $i$ \\
  $P_c\added[id=r2]{(\mathbf{x}, t)}$			& capillary pressure \\
  $\Psi_i\added[id=r2]{(\mathbf{x}, t)}$  & potential of fluid $i$ \\
  $\Psi_c\added[id=r2]{(\mathbf{x}, t)}$	& capillary potential \\
  $\mathbf{v}_i\added[id=r2]{(\mathbf{x}, t)}$			& Darcy's velocity of fluid $i$ \\
  $\mathbf{v}_t\added[id=r2]{(\mathbf{x}, t)}$			& total velocity \\
  $\mathbf{v}_c\added[id=r2]{(\mathbf{x}, t)}$     & capillary velocity \\
  $\rho_i$			 & density of fluid $i$ \\
  $\tilde{\mu}_i$			 & viscosity of fluid $i$ \\
  $m_i$  & mobility of fluid $i$: $m_i = k_{ri}/\tilde{\mu}_i$ \\
  $m_t$	 & total mobility: $m_t = m_{1} + m_{2}$ \\
  $q_i\added[id=r2]{(\mathbf{x}, t)}$			& injection or production rate of the $i$ fluid \\
  $g$				& gravity constant \\
  $Z$				& vector of the Z-direction \\
\hline
\end{tabular}
\label{tab:flow_physics_symbols}
\end{table}

The governing equations are derived from conservation of mass of each phase, and conservation of momentum or Darcy's law for each phase. First, we have
\begin{equation}\label{eqn:two_phase_mass}
      \frac{\partial }{{\partial t}}(\phi {S_i}{\rho _i}) + \nabla  \cdot ({\rho _i}{\mathbf{v}_i}) = {\rho _i}{q_i}, \quad i = 1,2
\end{equation}
where Eq.~(\ref{eqn:two_phase_mass}) is the mass balance equation, in which \(\phi\) denotes porosity, \(S_i\) denotes the saturation of the \(i\)-th phase, \(\rho_i\) denotes fluid density, \(\mathbf{v}_i\) denotes the volumetric velocity, and \(q_i\) stands for injection or production rates. The saturation of two phases should add up to 1, that is,
\begin{equation} \label{eqn:two_phase_sat}
      S_{1} + S_{2} = 1.
\end{equation}
In the remainder of the paper, we inject fluid 2 into the reservoir to replace fluid 1.

Darcy's law yields
\begin{equation}\label{eqn:two_phase_darcy}
      {\mathbf{v}_i} =  - \frac{{K{k_{ri}}}}{{{\tilde{\mu}_i}}}(\nabla {P_i} - g{\rho _i}\nabla Z), \quad i=1,2,
\end{equation}
where \(K\) is the permeability tensor, which is simplified as a scalar function of space in our model. \(\tilde{\mu}_i\) is the viscosity, \(P_i\) is the fluid pressure, \(g\) is the gravitational acceleration constant, and \(Z\) is the vector in the downward vertical direction. The relative permeability is written as \(k_{ri}\), which describes how easily one phase flows compared to the other. It is a nonlinear function of saturation of the corresponding phase, \textcolor{black}{such as the Corey model~\cite{corey1954interrelation}, the Brooks-Corey model~\cite{brooks1964hydrau}, and the van Genuchten-Mualem model~\cite{mualem1976new,van1980closed}}. \deleted[id=r0]{Usually, the higher the saturation, the easier the phase is to flow.}\textcolor{black}{Without loss of generality, we adopt a simplified Corey model for numerical experiments as follows
\begin{equation} \label{eqn:two_phase_relative_perm}
      k_{ri}(S_i) = S_i^2,
\end{equation}
where the residuals or irreducible saturations of the two phases are assumed to be zero and the end-point scaling factor is 1~\cite{lie2019introduction}. In practice, one may fit those parameters to measurements.} We define mobilities $m_i(S_i) = k_{ri}/\rho_i, i=1,2$, and the total mobility $m_t = m_1 + m_2$.
To complete the system, we use another equation to describe the relationship between the two phase-pressures as
\begin{equation}  \label{eqn:two_phase_capillary}
      P_{2} = P_{1} - P_c(S_{2}),
\end{equation}
where \(P_c\) is the capillary pressure, a function of the saturation of the wetting phase \(\alpha\), but it can be ignored and set to zero in fast-displacement processes.

In a nutshell, the flow physics model maps intrinsic properties to flow properties, for example,
\begin{equation}
      K(\mathbf{x}) \longrightarrow S_2(\mathbf{x},t).
\end{equation}We discuss the details of solving the flow equations in SI 1.1.

\subsection{Rock Physics Models}\label{sect:rock}
The second component in the physics system is the rock physics model that links flow properties to elastic properties. All symbols in the rock physics model are listed in \Cref{tab:rock_physics_symbols}. In our example, we construct the relationship between the saturation of the injected fluid $S_2$ to the rock bulk modulus $B_r$ and density $\rho_r$:
\begin{equation}
	B_r, \rho_r = \mathcal{R}(S_2; C_{p1}, C_{s1}, \rho_{r1} B_{f1}, B_{f2}, B_o, \phi, \rho_o, \rho_1, \rho_2),
\end{equation}
which are also controlled by the other elastic parameters in the parenthesis. Intuitively, the rock bulk modulus and density change with fluid substitution in the rock pore space, as the original and injected fluids may have different bulk moduli and densities. We first use the patchy saturation model that is commonly used for modeling of $\text{CO}_2$ injection processes~\cite{mavko2009rock}, and then examine a Gassmann's model with Brie's fluid mixing equation~\cite{brie1995shear,mavko2009rock}. In \replaced[id=r0]{both}{this} model\added[id=r0]{s}, we ignore the effects of pressure on the elastic properties. \added[id=r0]{One may find details about those two rock physics models in SI 1.2.}

\begin{table}[htpb]
\caption{Notation for the rock physics models. \added[id=r2]{The symbols $\mathbf{x}$ and $t$ denote space and time variables, respectively.}}
\centering
\begin{tabular}{l r}
\hline
 Symbol & Meaning  \\
\hline
  $B_{r1}\added[id=r2]{(\mathbf{x})}$  & bulk modulus of rock fully saturated with fluid 1   \\
  $B_{r2}\added[id=r2]{(\mathbf{x})}$  & bulk modulus of rock fully saturated with fluid 2   \\
  $B_{f1}$ & bulk modulus of fluid 1 \\
  $B_{f2}$ & bulk modulus of fluid 2 \\ 
  $B_{o}$  & bulk modulus of rock grains   \\
  $B_{f\text{\_mix}}(S_2)$ & bulk modulus of fluid mixture \\
  $\mu\added[id=r2]{(\mathbf{x})}$			 & rock shear modulus \\
  $C_{p1}\added[id=r2]{(\mathbf{x})}$ & P-wave velocity of rock fully saturated with fluid 1 \\
  $C_{p2}\added[id=r2]{(\mathbf{x})}$ & P-wave velocity of rock fully saturated with fluid 2 \\
  $C_{s1}\added[id=r2]{\mathbf{x})}$ & S-wave velocity of rock fully saturated with fluid 1 \\
  $C_{s2}\added[id=r2]{(\mathbf{x})}$ & S-wave velocity of rock fully saturated with fluid 2 \\
  $\phi\added[id=r2]{(\mathbf{x})}$	 & rock porosity \\
  $S_1\added[id=r2]{(\mathbf{x}, t)}$			 & saturation of fluid 1 \\
  $S_2\added[id=r2]{(\mathbf{x}, t)}$			 & saturation of fluid 2 \\
  $\rho_o$ & density of rock grains\\
  $\rho_{1}$ & density of fluid 1\\
  $\rho_{2}$ & density of fluid 2\\
  $\rho_{r1}$ & density of rock fully saturated with fluid 1\\
  $\rho_r\added[id=r2]{\left(S_2\left(\mathbf{x}, t\right)\right)}$ & density of rock as a function of saturation of fluid 2\\
  $B_r\added[id=r2]{\left(S_2\right(\mathbf{x}, t\left)\right)}$ & bulk modulus of rock as a function of saturation of fluid 2\\
\hline
\end{tabular}
\label{tab:rock_physics_symbols}
\end{table}

\deleted[id=r0]{In this paper, we consider two rock physics models: the patchy saturation model and the Gassmann's model with Brie's fluid mixture equations, whose details are presented in \mbox{\Cref{subsect:rock_physics_more}}.}

\subsection{The Elastic Wave Equation}\label{sect:elastic}
The wave phenomenon can be described by the following velocity-stress formulation of the elastic wave equation:
\begin{eqnarray} \label{eqn:elastic_wave}
	&& \rho \frac{\partial v_i}{\partial t} = \frac{\partial \sigma_{ij}}{\partial x_j} + f_i  \nonumber \\
	&& \frac{\partial \sigma_{ij}}{\partial t} = \lambda \frac{\partial v_k}{\partial x_k}\delta_{ij} + \mu\left( \frac{\partial v_i}{\partial x_j} + \frac{\partial v_j}{\partial x_i} \right),
\end{eqnarray}
where $v_i$ is the particle velocity vector, $\sigma_{ij}$ is the stress tensor, $f_i$ is the body force, and the repeated index indicates the Einstein summation notation. Density is denoted by $\rho$ and the Lam\'e parameters are represented by $\lambda$ and $\mu$ (shear modulus). The first Lam\'e parameter \(\lambda\) and bulk modulus \(B\) that appears in the rock physics model are connected by
\begin{equation}
      \lambda = B - \frac{2}{3}\mu.
\end{equation}
We assume that the depth of the studied area is deep enough so that free surface reflected waves do not appear in the recorded data, and we use convolution perfectly matched layers (CPML)~\cite{martin2008unsplit} on the four boundaries to prevent unrealistic reflected waves.

In summary, the wave physics model maps elastic properties to wavefields:
\begin{equation}
      B(\mathbf{x}), \,\mu(\mathbf{x}), \,\rho(\mathbf{x}) \longrightarrow \sigma_{zz}(\mathbf{x},t), \, \sigma_{xx}(\mathbf{x},t), \, \sigma_{xz}(\mathbf{x},t), \, v_z(\mathbf{x},t), \, v_x(\mathbf{x},t).
\end{equation}

\section{Inversion Method -- \replaced{Intrusive}{Intelligent} Automatic Differentiation}\label{sect:inversion_method}
Automatic differentiation has been successfully used in scientific computing and is one of the major contributors to the success and accessibility of deep-learning technology. Within modern deep-learning frameworks such as \texttt{TensorFlow}~\cite{abadi2016tensorflow}, there is a suite of built-in differentiable operators (or neural network layers). With those operators, users may construct complex forward mapping functions and automatically obtain gradients. Besides, those frameworks offer high-performance computing options, such as GPU acceleration and graph-based parallelization. Those features are highly desirable when implementing complicated functions such as rock physics models. However, it is computationally inefficient to solve PDEs with these built-in operators or layers. Also, automatic differentiation needs to store intermediate variables for gradient back-propagation, which is also demanding for memory resources.

On the other hand, the so-called adjoint method is widely used for large-scale PDE-constrained inverse problems, such as full-waveform inversion, history matching~\cite{li2001history,oliver2011recent}, 4D variational weather data assimilation~\cite{wang2001review}, ocean model inversion~\cite{marotzke1999construction}, etc. The adjoint method requires analytic derivations of the adjoint equations and gradients. Here we follow the ``discretize-then-optimize'' philosophy, which means that one first discretizes the forward equations and then makes the following derivations according to the Lagrange multiplier method. We review the details of the discrete adjoint method in SI 2 and illustrate its application to elastic FWI in SI 4. Once the adjoint equations and formulae of gradients are obtained, one may use various techniques to optimize computation efficiency and reduce the memory footprint. For example, it is not possible to store the whole history of forward wave propagation, so we only store wavefields at the boundaries, and reverse the forward wave equations in time to recompute the forward wavefields. The users can adopt other memory-saving strategies, such as saving a few frequency components~\cite{zhang2018hybrid} or using check-pointing methods~\cite{anderson2012time}. It is possible to apply the discrete adjoint method to the whole subsurface flow FWI problem, but the derivations are time-consuming, error-prone, and inflexible, especially when dealing with systems where there are complicated time-stepping schemes and mappings between variables. After all, the derivations are purely mechanical and should be automated.

Therefore, we propose an \replaced{intrusive}{intelligent} automatic differentiation strategy that provides different levels of customization. It combines native automatic differentiation and the discrete adjoint method, such that we can achieve a good balance between computational efficiency and implementation efficiency or flexibility. This is realized by implementing customized PDE operators with gradient computation capabilities that resemble neural network layers in deep-learning frameworks. 

As we show in SI 3, automatic differentiation and discrete adjoint method are equivalent. The only difference is that automatic differentiation requires a specific input-output interface in which we only need to implement two methods for each customized operator:
\begin{enumerate}
	\item Forward operation $f_n$: compute the output by solving a PDE:
            \begin{equation}
                  \mathbf{u}_{n+1} = f_n(\mathbf{u}_n, \bm{\theta}),
            \end{equation}
            \replaced[id=r2]{For example, $f_n$ represents a single-step discrete-time propagation of a wave equation, $\mathbf{u}_n$ is the wavefield (a vector of particle velocity and stress) at the $n$-th time step, and $\bm{\theta}$ is the Lam\'e parameters. In the general case, $\mathbf{u}_n$ can be any state variables of a PDE, and $\bm{\theta}$ is the PDE parameter, which depends on the physical quantity we want to calibrate.}{where \(\mathbf{u}_n\) is a physical quantity that the PDE describes, such as fluid saturation, pressure in flow equations or wavefield (e.g., particle velocity, stress) in wave equations at time step \(n\), and \(\bm{\theta}\) is the parameter of the PDE. The forward-modeling operator \(f_n\) symbolically represents the numerical solving of the PDE at time step \(n\).}
            
	\item Backward operation $b_n$: given gradient $\frac{\partial \mathcal{J}}{\partial \mathbf{u}_{n+1}}$, compute $\frac{\partial \mathcal{J}}{\partial \mathbf{u}_{n+1}} \frac{\partial f_n}{\partial \mathbf{u}_{n}}$ and $\frac{\partial \mathcal{J}}{\partial \mathbf{u}_{n+1}}\frac{\partial f_n}{\partial \theta}$ with the chain rule. The backward process \replaced[id=r1]{back-propagates}{sprays} the gradient of the objective function with respect to the output to the input variables:
		\begin{equation}
		  \frac{\partial \mathcal{J}}{\partial \mathbf{u}_{n+1}}\frac{\partial f_n}{\partial \mathbf{u}_{n}}, \quad \frac{\partial \mathcal{J}}{\partial \mathbf{u}_{n+1}}\frac{\partial f_n}{\partial \bm{\theta}}= b_n\left(\frac{\partial \mathcal{J}}{\partial \mathbf{u}_{n+1}}, \mathbf{u}_{n+1}, \mathbf{u}_n, \bm{\theta}\right).
		\end{equation}
\end{enumerate}
Note that only the matrix-vector product needs to be implemented, and there should be no need to construct Jacobian matrices like $\frac{\partial f_n}{\partial \mathbf{u}_n}$ explicitly.

The customized PDE operators enable two different levels of customization. For example, 
\begin{enumerate}
\item the forward process can be wave propagation plus misfit computation, and the backward gradient is computed by the adjoint method. In other words, we encapsulate the entire FWI in an operator.
\item the forward process can be a forward step in the numerical solving of PDEs. This level of customization is helpful when the forward step is numerically complicated. For example, in our flow simulations, we solve nonlinear equations with the Newton-Raphson method and use implicit-time stepping.
\end{enumerate}

\replaced[id=r2]{Specifically}{To be more specific about the second level of customization}, for explicit time-stepping, the implementation of \texttt{backward} is straight-forward, as one can usually derive it from the forward code directly. However, we need to employ special tricks for the implicit time-stepping. Now we consider a problem with implicit constraints:
\begin{equation}
 \begin{aligned}
     \min_{\mathbf{u}_1, \bm{\theta}} &\  f_4(\mathbf{u}_1, \mathbf{u}_2, \mathbf{u}_3, \mathbf{u}_4), \\
     \mathrm{s.t.} & \ g_1(\mathbf{u}_2, \mathbf{u}_1, \bm {\theta}) = 0, \\
     & \ g_2(\mathbf{u}_3, \mathbf{u}_2, \bm {\theta}) = 0,\\
     & \ g_3(\mathbf{u}_4, \mathbf{u}_3, \bm {\theta}) = 0,
\end{aligned}
\end{equation}
where the set of equations $g_n(\mathbf{u}_{n+1}, \mathbf{u}_n, \bm {\theta})=0\, (n = 1,2,3)$ establish the forward mappings $\mathbf{u}_{n+1} = f_n(\mathbf{u}_n, \bm {\theta}),\, (n = 1,2,3)$, respectively.

However, the programming interface for \texttt{backward} is still to compute
\begin{equation} 
\label{eqn:Du_Du}
     \frac{\partial \mathcal{J}}{\partial \mathbf{u}_{n+1}}\frac{\partial f_n}{\partial  \mathbf{u}_{n}},
\end{equation}
when we are supplied with \(\frac{\partial \mathcal{J}}{\partial \mathbf{u}_{n+1}}\).

We compute them using the following technique:
Take the partial derivative with respect to the interested parameter (e.g., \(\mathbf{u}_n\)) on the constraint equation $g_i(\mathbf{u}_{n+1}, \mathbf{u}_n, \bm {\theta}) = 0$ and obtain
\begin{equation}
 \frac{\partial g_{n}}{\partial \mathbf{u}_{n+1}}\frac{\partial f_n}{\partial \mathbf{u}_{n}} + \frac{\partial g_{n}}{\partial \mathbf{u}_{n}} = 0,
\end{equation}
and thus
\begin{equation}
 \frac{\partial f_n}{\partial \mathbf{u}_{n}} = -\left( \frac{\partial g_{n}}{\partial \mathbf{u}_{n+1}} \right) ^{-1} \frac{\partial g_{n}}{\partial \mathbf{u}_{n}}.
\end{equation}
Therefore,
\begin{equation}\label{equ:deriv}
     \frac{\partial \mathcal{J}}{\partial \mathbf{u}_{n+1}}\frac{\partial f_n}{\partial \mathbf{u}_{n}} = -\frac{\partial \mathcal{J}}{\partial \mathbf{u}_{n+1}} \left( \frac{\partial g_{n}}{\partial \mathbf{u}_{n+1}} \right) ^{-1} \frac{\partial g_{n}}{\partial \mathbf{u}_{n}}.
\end{equation}
If we let
\begin{equation}
     \bm{\lambda}_{n+1}^T = -\frac{\partial \mathcal{J}}{\partial \mathbf{u}_{n+1}} \left( \frac{\partial g_n}{\partial \mathbf{u}_{n+1}} \right) ^{-1},
\end{equation}
then we can get \(\bm{\lambda}_{n+1}\) by solving the following linear system
\begin{equation}
\label{eqn:ad_implicit_lambda}
     \left( \frac{\partial g_n}{\partial \mathbf{u}_{n+1}} \right)^T \bm{\lambda}_{n+1} = -\left(\frac{\partial \mathcal{J}}{\partial \mathbf{u}_{n+1}}\right)^T,
\end{equation}
and then can get
\begin{equation}
\label{eqn:ad_implicit_gradient}
     \frac{\partial \mathcal{J}}{\partial \mathbf{u}_{n+1}}\frac{\partial f_n}{\partial  \mathbf{u}_{n}} = \bm{\lambda}_{n+1}^T \frac{\partial g_n}{\partial \mathbf{u}_n}.
\end{equation}
To summarize, we (1) solve linear system Eq.~(\ref{eqn:ad_implicit_lambda}) for adjoint variables $\bm {\lambda}_i$, and (2) compute the matrix-vector product as in Eq.~(\ref{eqn:ad_implicit_gradient}).

\begin{example}

\textcolor{black}{Consider the following constrained optimization problem
\begin{align}
    \min_{\theta} &\; \frac{1}{2}\|\mathbf{u} - \mathbf{u}_{\mathrm{obs}}\|^2\\
    \mathrm{s.t.} & \; A(\theta)\mathbf{u} = \mathbf{b}
\end{align}
where $\theta$ is a scalar parameter to be calibrated, $\mathbf{u}$ is the state variable of a PDE whose discrete form is given by $A(\theta)\mathbf{u} = \mathbf{b}$, $\mathbf{u}_{\mathrm{obs}}$ is the observation vector, $\mathbf{b}$ is a known source term, and $A(\theta)$ is the coefficient matrix whose entries depend on $\theta$. Our goal is to compute the derivative of $$\mathcal{J}(\theta) := \frac{1}{2}\|\mathbf{u}(\theta) - \mathbf{u}_{\mathrm{obs}}\|^2$$
where $\mathbf{u}(\theta)$ is the solution to the constraint $ A(\theta)\mathbf{u} = \mathbf{b}$, i.e., $\mathbf{u}(\theta) = A(\theta)^{-1} \mathbf{b}$. }

\textcolor{black}{
Let 
$$g(\theta, \mathbf{u}) = A(\theta)\bm{u} - \mathbf{b}$$
Using \Cref{equ:deriv}, we have (subtitute $\mathbf{u}_{n+1}$ with $\mathbf{u}$, $\mathbf{u}_n $ with $\theta$, and $g_n$ with $g$)
\begin{align*}
\frac{\partial \mathcal J}{\partial \theta} &= -\frac{\partial \mathcal J}{\partial \mathbf{u}} \left( \frac{\partial g}{\partial \mathbf{u}} \right)^{-1} \frac{\partial g}{\partial {\theta}}    \\
& =  (\mathbf{u}(\theta) - \mathbf{u}_{\mathrm{obs}})^T A(\theta)^{-1} \frac{\partial A(\theta)}{\partial \theta} \mathbf{u} 
\end{align*}
In the essence of \Cref{eqn:ad_implicit_lambda}, we first calculate $\bx^T = (\mathbf{u}(\theta) - \mathbf{u}_{\mathrm{obs}})^T A(\theta)^{-1}$ by solving the linear system 
$$A(\theta)^T \bx = \mathbf{u}(\theta) - \mathbf{u}_{\mathrm{obs}}$$
and finally we have 
$$\frac{\partial \mathcal J}{\partial \theta}  = \bx^T \frac{\partial A(\theta)}{\partial \theta} \mathbf{u} $$
In the intrusive automatic differentiation, this process can be incorporated into an automatic differentiation system and therefore exchanges state variable and gradient information with other physical modules. For more implementation details, see our software FwiFlow.jl and the supporting information (SI) material. 
}
\end{example}

\section{Numerical Experiments}\label{sect:numerical_experiments}
\subsection{Coupled vs. Decoupled Inversion}
Our examples are inspired by $\text{CO}_2$ sequestration projects for the reduction of $\text{CO}_2$ emission into the atmosphere. In this example, an injection well injects super-critical $\text{CO}_2$ at a constant rate. It would be more accurate to implement a constant-flow-rate boundary condition to represent an open aquifer, but for simplicity we put another well that produces water displaced by $\text{CO}_2$ as an artificial approximation.

The flow physics can be approximated by immiscible incompressible two-phase flow, even though the model can be much more complicated. As the saturation of $\text{CO}_2$ increases, the bulk modulus and density of the saturated rocks change according to certain rock physics models. These changes in elastic properties can be sensed by elastic waves.

\begin{figure}[htpb]
	\begin{center}
		\setlength{\tabcolsep}{0.2cm}
		\begin{tabular}{l l}
			\small{(a)} & \small{(b)}\\
			\includegraphics[height=5cm]{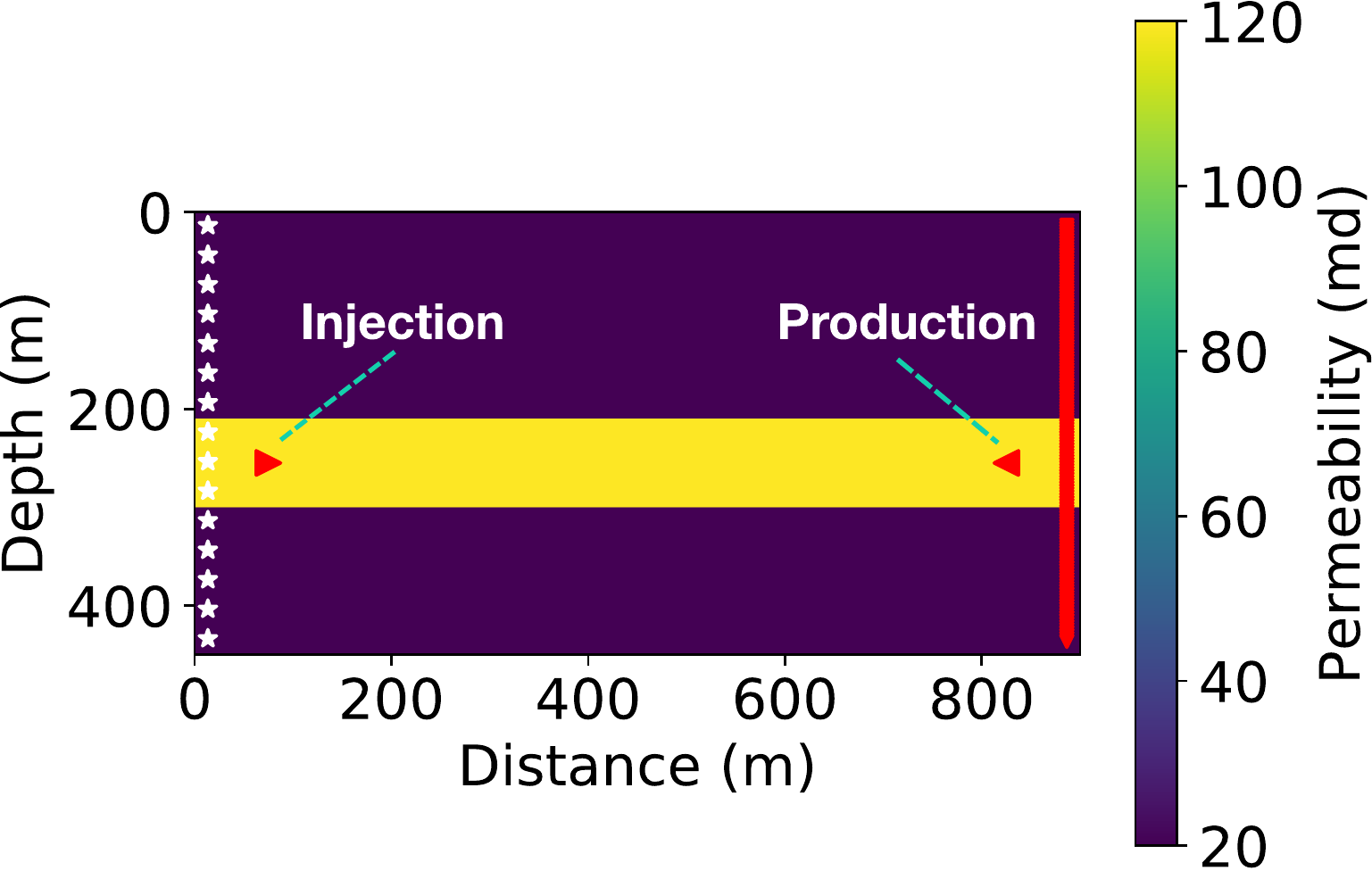}
			&  \includegraphics[height=5cm]{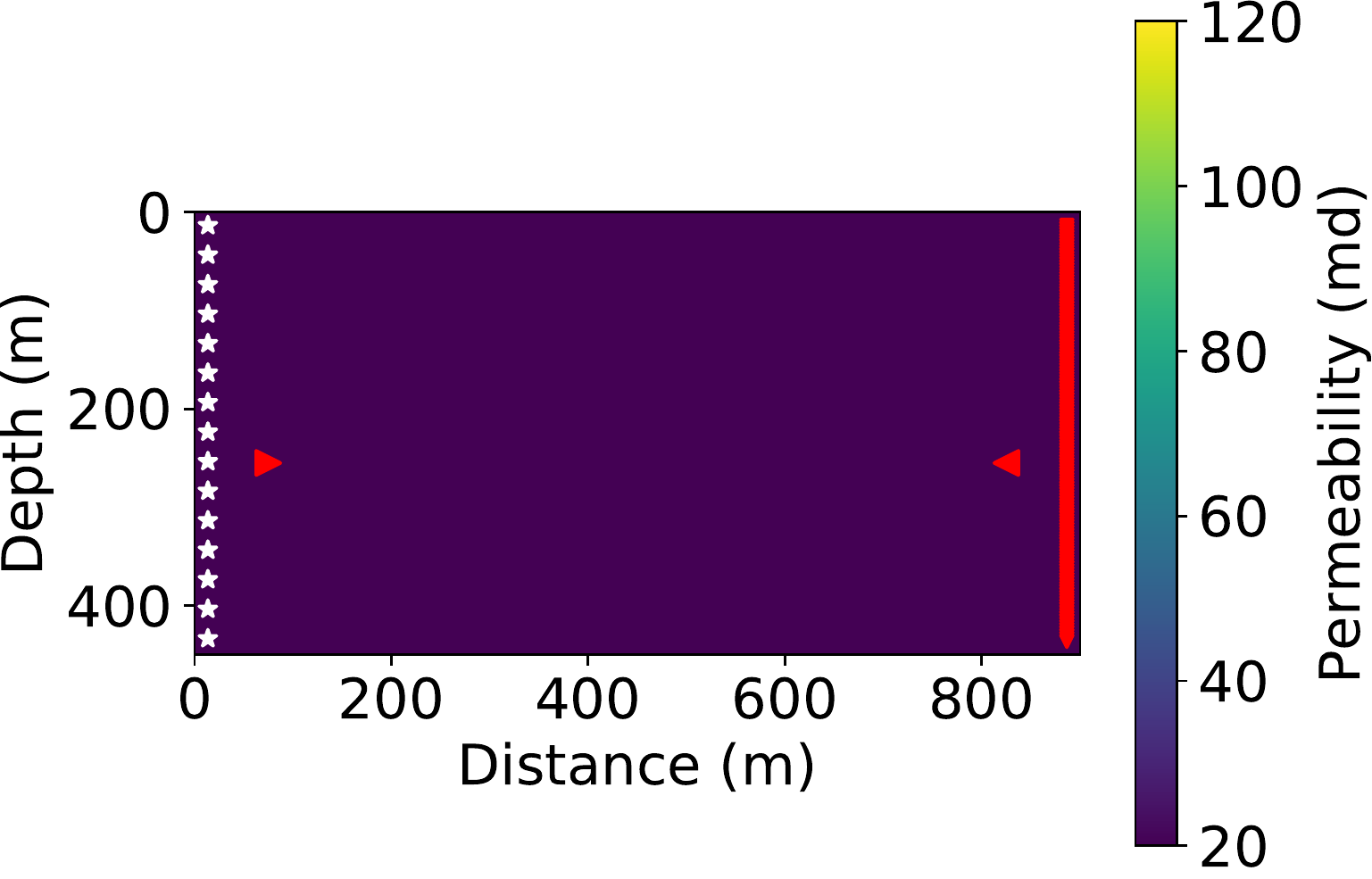}
		\end{tabular}
	\end{center}
	\caption{True and initial permeability models. \added[id=r3]{The red arrowheads stand for the injection or production locations, the white stars denote the seismic source locations, and the red lines denote the seismic receiver locations.}}
	\label{fig:K_true_init}
\end{figure}

The dimension of the permeability models is $450\times 900 \times 10$~m, with 15 cells in the depth direction, 30 cells in the horizontal direction. Each cell is a $30\times30\times10$~m block. In the true model, the background permeability is 20 millidarcy (md), while there is a high-permeability layer of 120 md in the middle, as shown in Fig.~\ref{fig:K_true_init}. In the initial model, the permeability in the whole model is 20 md. Before injection, the reservoir was filled with water, whose density is 1053.0~$\mathrm{kg}/\mathrm{m}^3$ and viscosity is 1.0~centipoise ($\mathrm{cP}$). The injected supercritical $\text{CO}_2$ has a density of 501.9~$\mathrm{kg}/\mathrm{m}^3$ and a viscosity of 0.1~$\mathrm{cP}$. The water injection rate was 0.005~$\mathrm{m}^3/\mathrm{s}$ and the production well produces water at a rate of 0.005~$\mathrm{m}^3/\mathrm{s}$ as well. \added[id=r2]{The initial $V_p$, $V_s$, and density models before injection are of constant values, which are 3500 m/s, $3500/\sqrt{3}$ m/s, and 2200 kg/m${}^3$, respectively.} We simulated 51 slow-time steps at intervals of 20 days, which gives 1000-day total simulation time. Eleven of the 51, including the initial state, are used for slow-time measurements.

\begin{figure}[htpb]
\noindent\includegraphics[width=\textwidth]{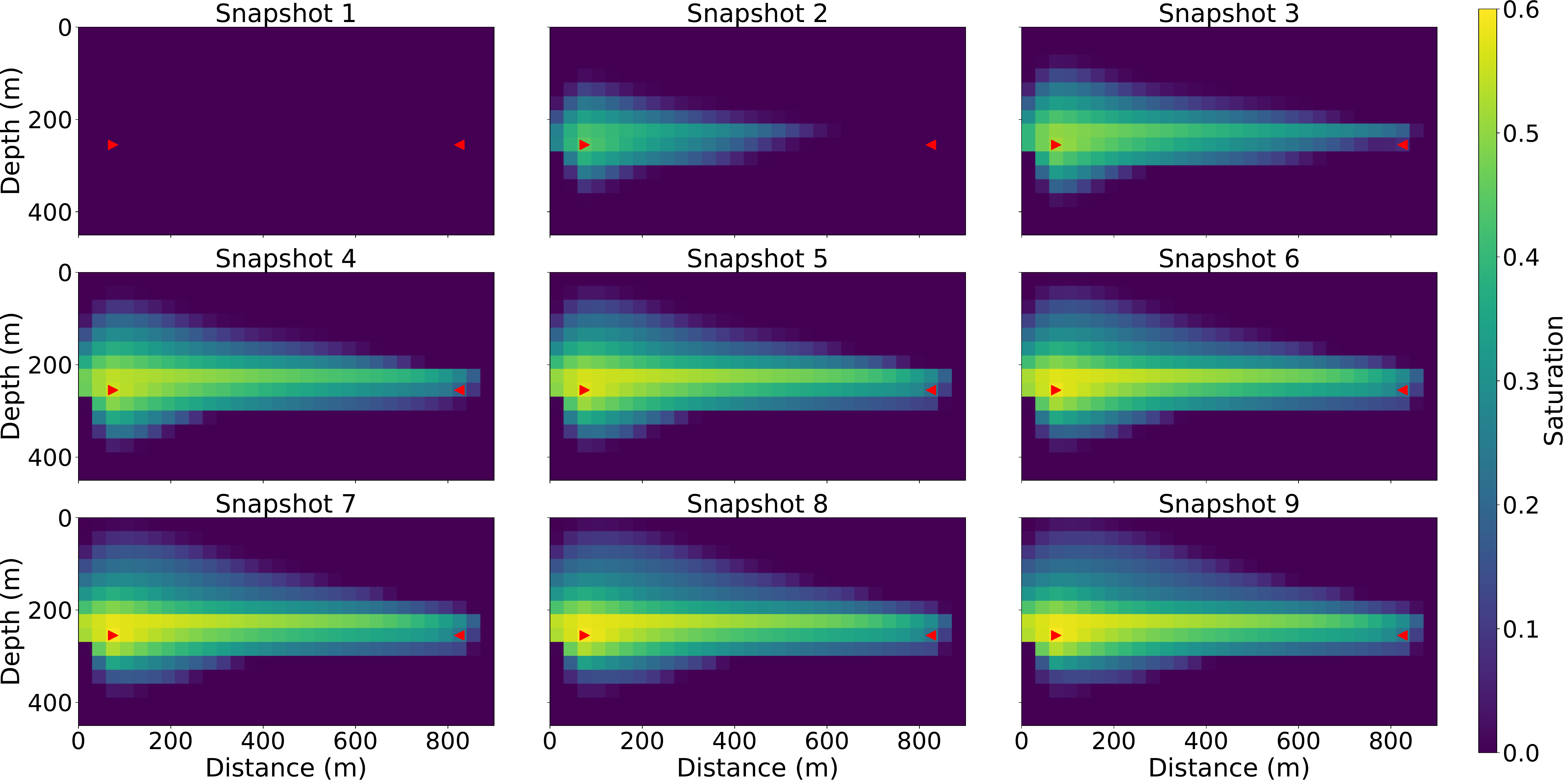}
\caption{Evolution of $\text{CO}_2$ saturation of the first 9 stages with the true permeability model.}
\label{fig:Sat_evo_patchy_true}
\end{figure}

Of course the $\text{CO}_2$ plume favors the highly permeable layer (Fig.~\ref{fig:Sat_evo_patchy_true}) and migrate from the injection well (red triangle on the left) to the production well (red triangle on the right), while if $\text{CO}_2$ is injected in a homogeneous reservoir, the plume expands more symmetrically towards the production well with slightly upward movement due to buoyancy.

\begin{figure}[htpb]
\noindent\includegraphics[width=\textwidth]{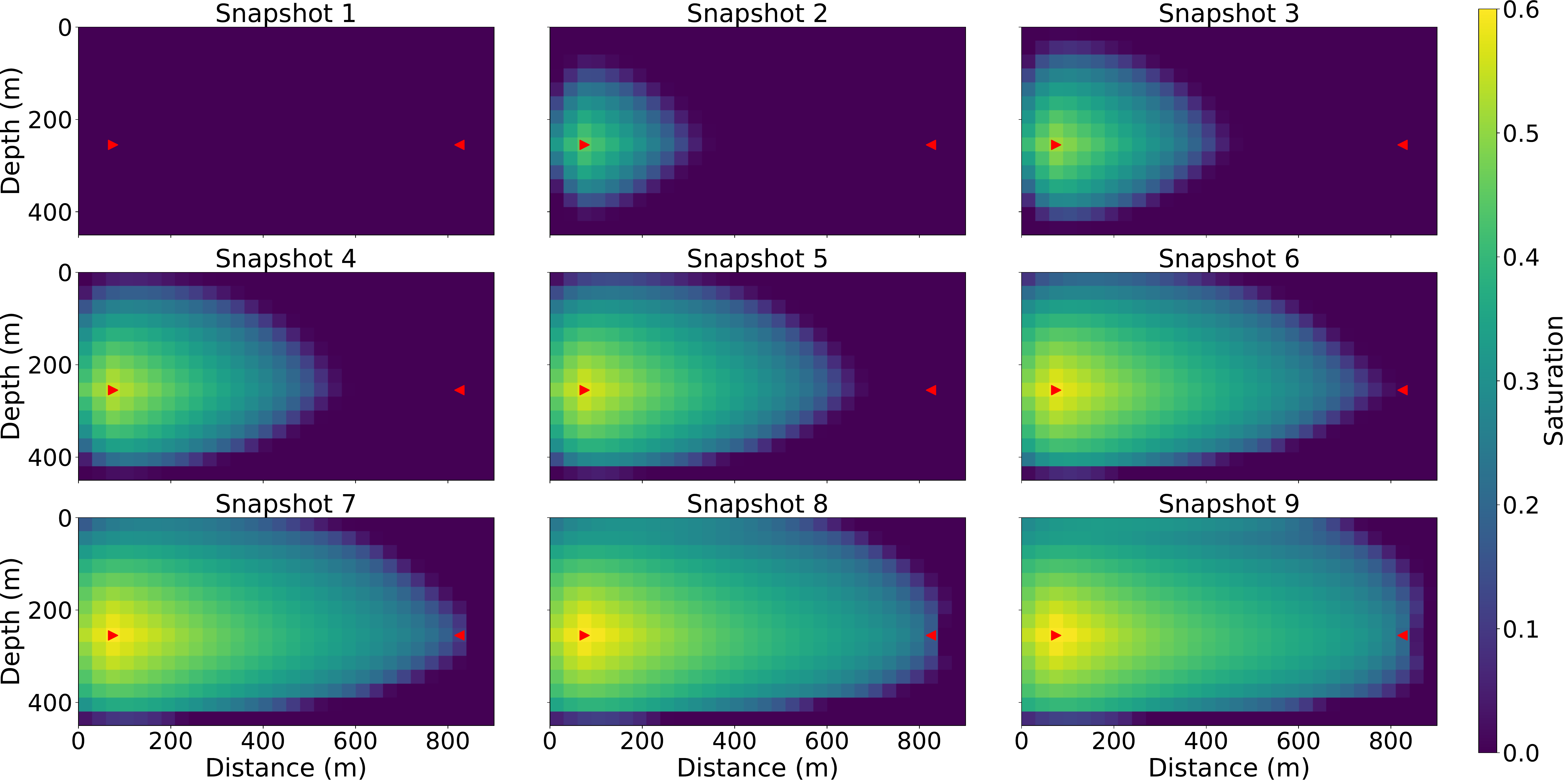}
\caption{Evolution of $\text{CO}_2$ saturation of the first 9 stages with the initial permeability model.}
\label{fig:Sat_evo_patchy_init}
\end{figure}

We up-sampled the saturation evolution maps to a spatial grid interval of 3~$\mathrm{m}$, while keeping the vertical and horizontal dimensions the same. Then, the saturation maps were transformed to elastic properties, Lam\'e parameter $\lambda$, $\mu$ and density $\rho$ in our case, with the patchy saturation rock physics model.

\begin{figure}[htpb]
	\begin{center}
		\setlength{\tabcolsep}{0.2cm}
		\begin{tabular}{l l l}
			\small{(a)} & \small{(b)} & \small{(c)}\\
			\includegraphics[height=6cm]{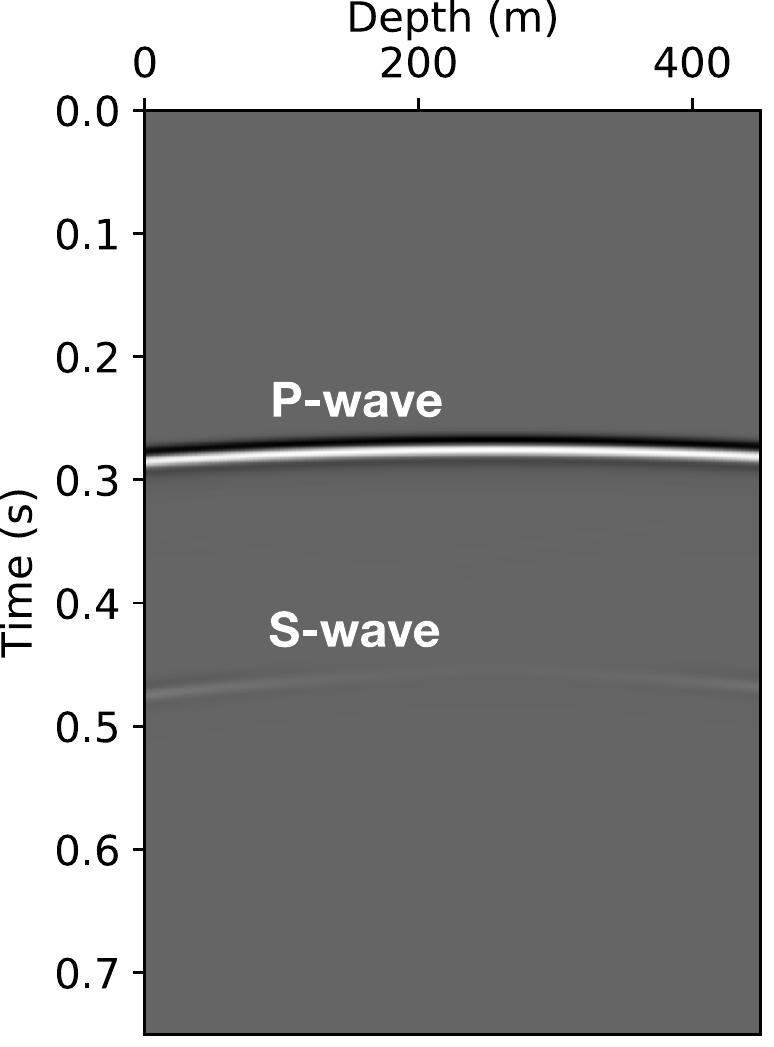}
			&  \includegraphics[height=6cm]{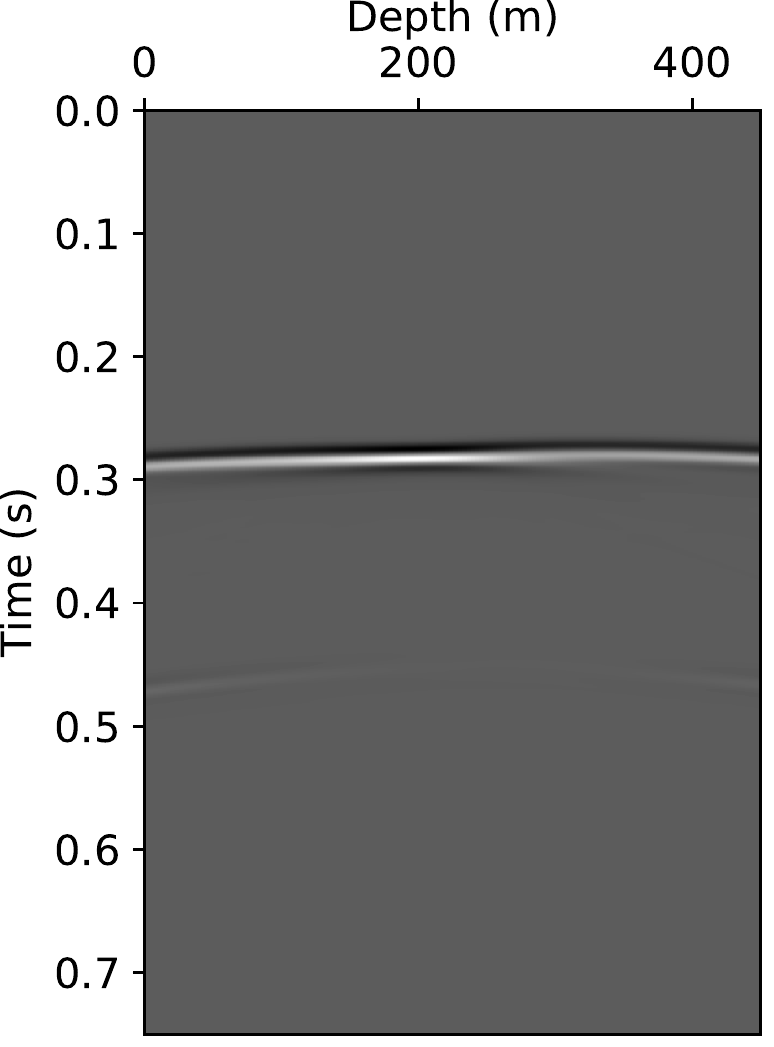} & \includegraphics[height=6cm]{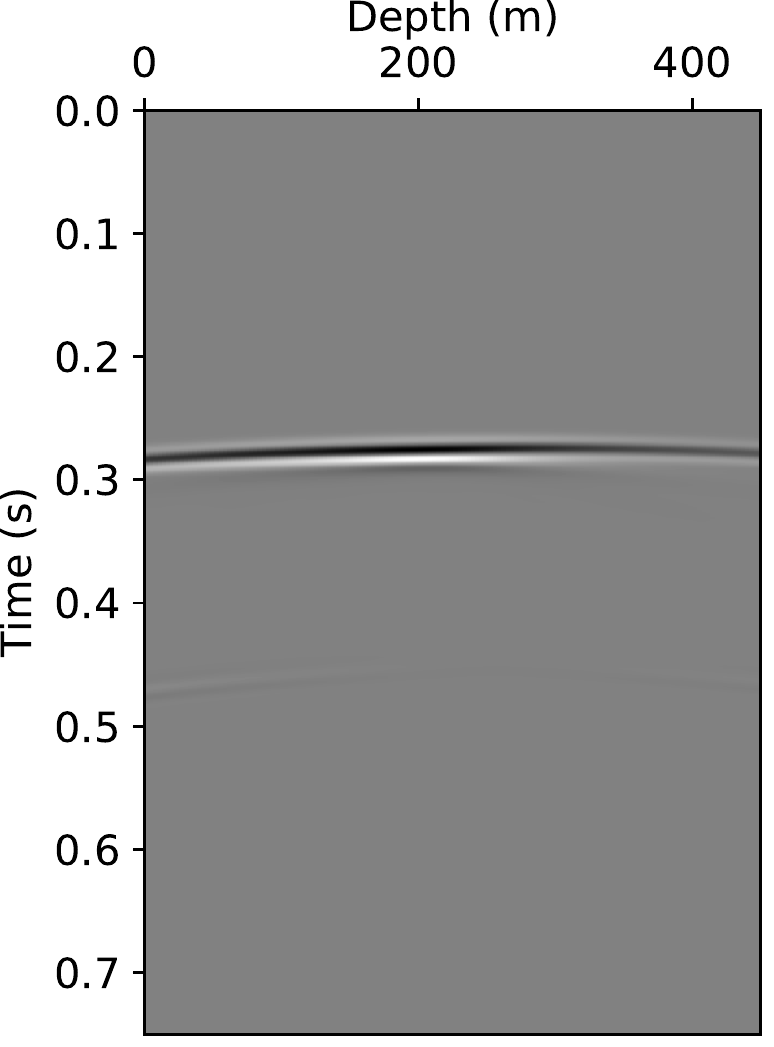}
		\end{tabular}
	\end{center}
	\caption{Seismic shot gathers of the 8th source. (a) Survey 1, (b) survey 11, (c) difference between (a) and (b).}
	\label{fig:seismic_data}
\end{figure}

We conducted 11 crosswell seismic surveys to measure the elastic property evolutions with the slow time. In the base case, there were 15 seismic sources and 142 receivers separated by 873~$\mathrm{m}$. \textcolor{black}{We adopted a 2-D tensile crack model~\cite{aki2002quantitative} to approximate a specific radiation/sensitivity pattern of borehole sources and receivers (SI 8).} The source time function was a Ricker wavelet with a dominant frequency of 50~$\mathrm{Hz}$. The seismic simulation time step was 0.00025~$\mathrm{s}$, and the whole simulation time was 0.75~$\mathrm{s}$. The shot gathers of the 8th source at the first and the eleventh survey are shown in Fig.~\ref{fig:seismic_data}. In each shot gather, the prominent events are a strong P-wave followed by a weaker S-wave. We see a slight delay in arrival time from the 1st to 11th survey, as well as slight changes in amplitude and wavefront shape in the last survey.

The optimization problem was solved by the L-BFGS-B method~\cite{byrd1995limited} with a box constraint on the permeability value between 10~$\mathrm{md}$ and 130~$\mathrm{md}$. \added[id=r0]{L-BFGS-B is short for limited memory Broyden-Fletcher-Goldfarb-Shanno algorithm with box constraints, a standard variant of a BFGS method, which is an efficient and popular quasi-Newton algorithm for nonsmooth (both convex and nonconvex) optimization. }\textcolor{black}{In this work, we use the line search routine in \cite{more1994line}, which attempts to enforce the Wolfe conditions \cite{byrd1995limited} by a sequence of polynomial interpolations.}

In the coupled inverted model (Fig.~\ref{fig:Inversion_result_sep_fit}(a)), the high permeability layer was well constructed, though there are minor fluctuations in the value within the layer. To quantify the inversion accuracy, we define the mean-squared-error (MSE) measure between the true and the inverted permeability models as
\begin{equation}
	\text{MSE} = \frac{1}{N_z}\frac{1}{N_x}\sum_{i=1}^{N_z}\sum_{j=1}^{N_x}(K^{\text{true}}_{ij} - K^{\text{inv}}_{ij})^2,
\end{equation}
where $N_z$ and $N_x$ are the two dimensions of discretized permeability models. We found that the MSE of the model from coupled inversion is 218.71.

\begin{figure}[htpb]
\noindent\includegraphics[width=\textwidth]{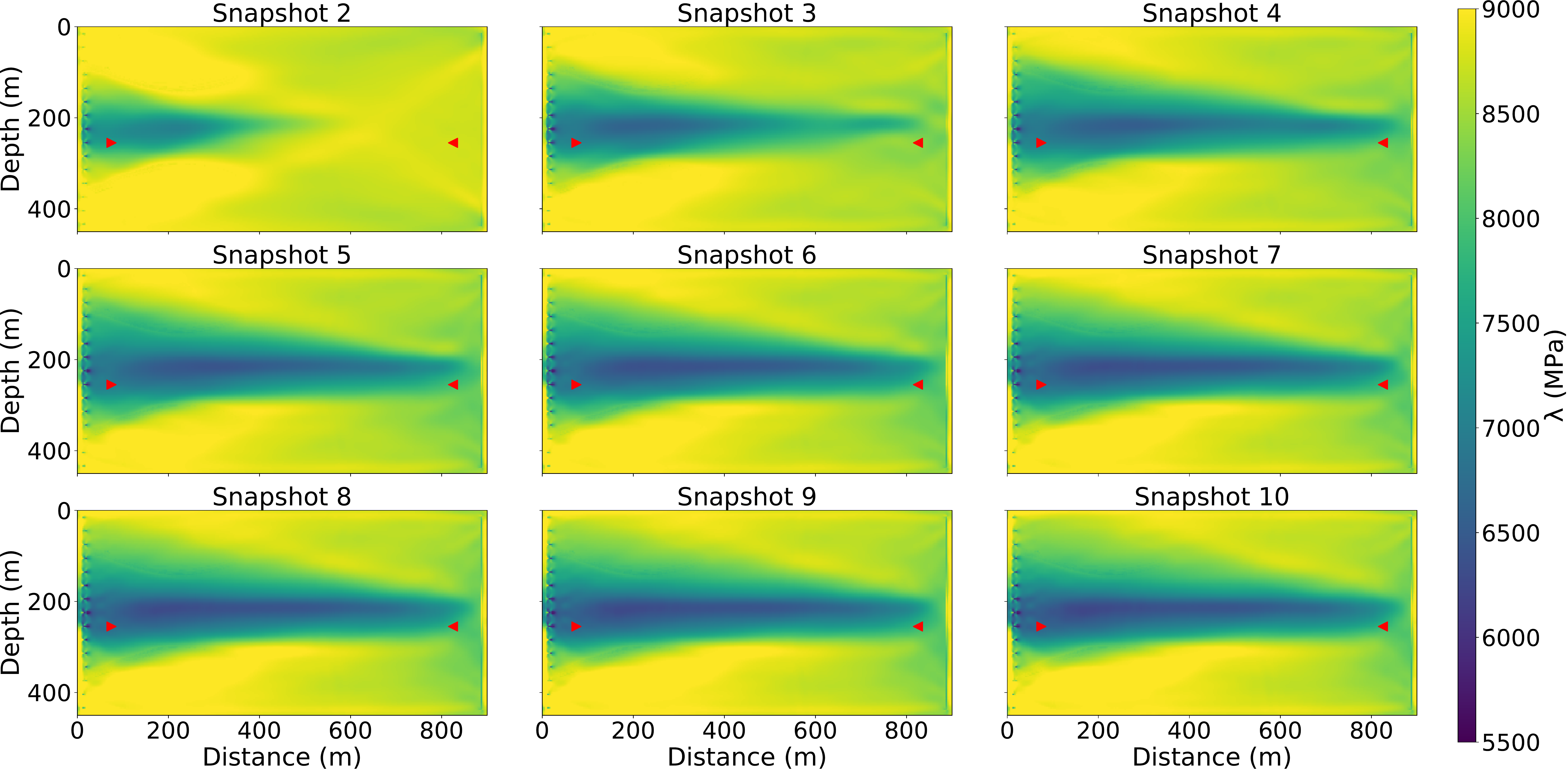}
\caption{Inverted Lam\'e parameter $\lambda$ for nine different surveys. The inversion is conducted using FWI with the same data separately for each survey. The survey configurations are the same as the baseline case.}
\label{fig:lambda_inverted}
\end{figure}

\begin{figure}[htpb]
\noindent\includegraphics[width=\textwidth]{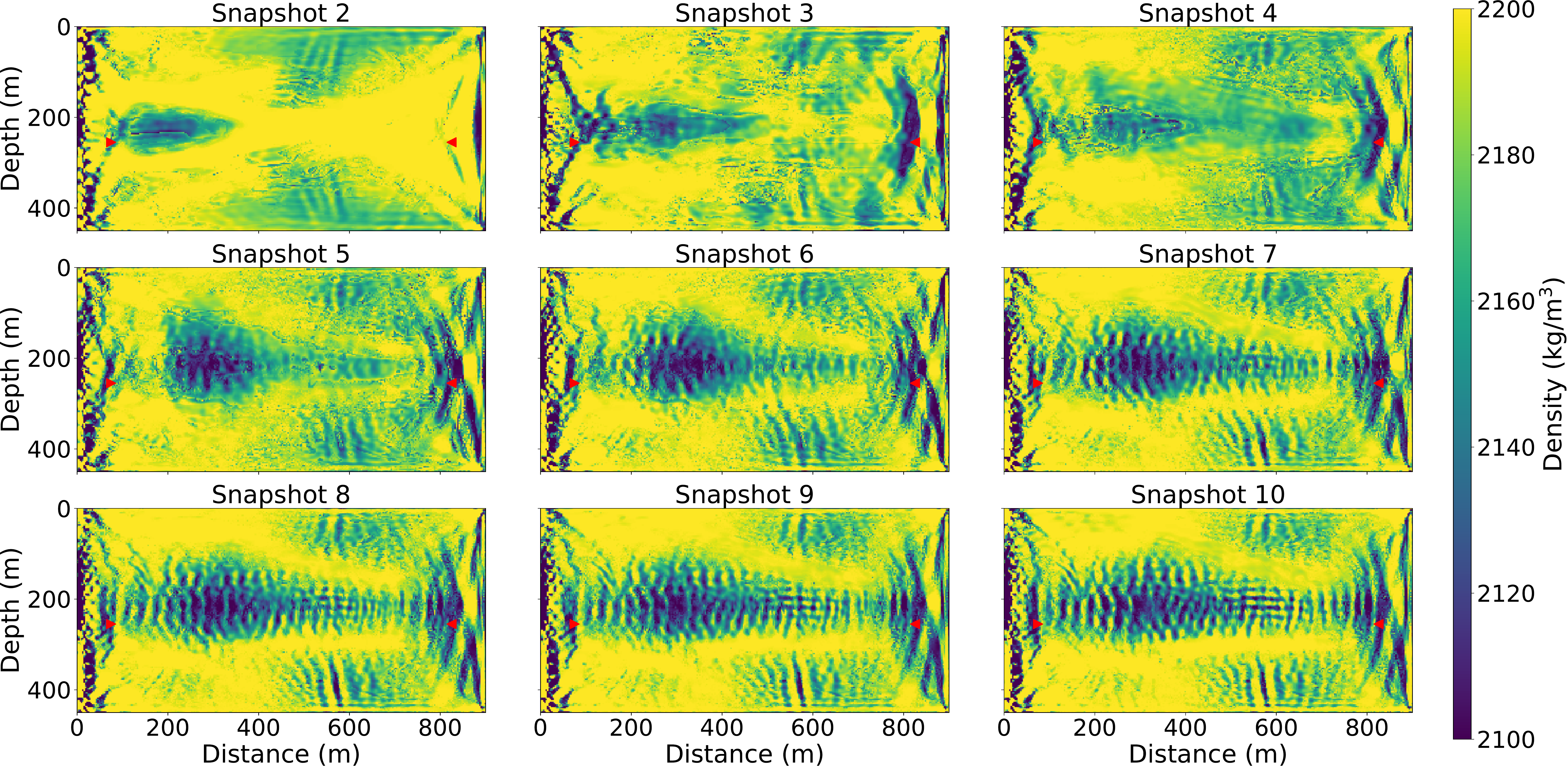}
\caption{\added[id=r2]{Inverted density for nine different surveys. The inversion is conducted using FWI with the same data separately for each survey. The survey configurations are the same as the baseline case. The results are contaminated with strong artifacts.}}
\label{fig:den_sep_inverted}
\end{figure}

\begin{figure}[htpb]
\noindent\includegraphics[width=\textwidth]{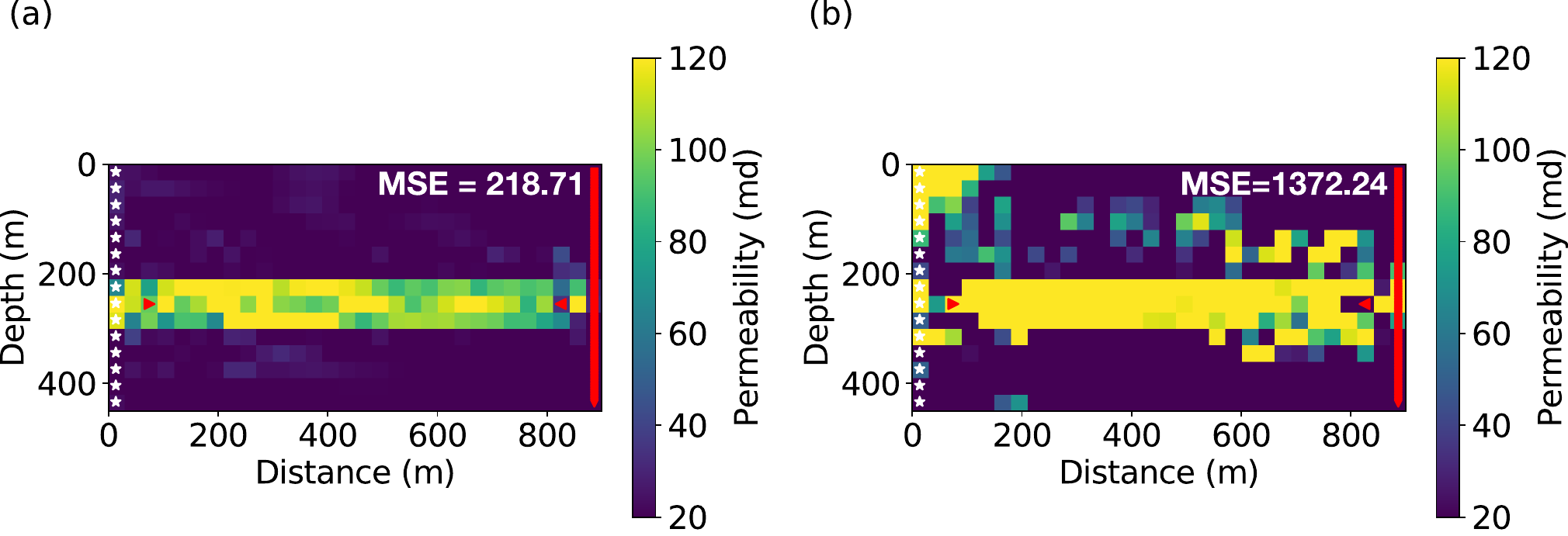}
\caption{Comparison of inverted permeability model using (a) coupled inversion (baseline, $\text{MSE}=218.71$), and (b) using $\lambda$ separately inverted as shown in Fig.~\ref{fig:lambda_inverted} ($\text{MSE}=1372.24$). The separately fitted permeability model has much stronger artifacts and much larger MSE.}
\label{fig:Inversion_result_sep_fit}
\end{figure}

There exists a traditional choice for permeability field inversion, which we term \replaced{\textbf{decoupled}}{\texttt{decoupled}} inversion: one may separately invert for elastic properties at each snapshot using FWI, and then use the inverted elastic properties as data to fit the flow equation to reconstruct the hidden intrinsic parameter field, similar to the seismic history matching methods that we described in the introduction. \added[id=r2]{This strategy has its merits of being easy to implement and helpful in quality control of each inversion component.}\deleted{This strategy is relatively easier to implement, but ignores the slow-time dynamics of the coupled inversion process. The inverted results for each survey may have artifacts from various sources. If they are treated as observed data to fit the slow-time process, then we expect that the final inversion result would exhibit stronger artifacts than that from the coupled inversion.} \added[id=r0]{However, seismic inversion usually carries significant inversion artifacts due to various reasons, such as limited aperture, noises in the data, and being trapped in local minima. Such seismic artifacts cannot be predicted by flow physics, and thus would lead to artifacts in the inverted intrinsic parameters in the decoupled inversion. In the coupled inversion, the flow physics PDE plays the role of spatial-temporal model regularization, which gets rid of the intermediate step in the decoupled inversion where seismic inversion artifacts accumulate. From another perspective, the coupling reduces independent elastic parameters from multiple surveys to the intrinsic parameters that do not change with time. The dimension reduction and inversion in the latent space (the space of intrinsic parameters) contribute to the reduction of artifacts.} We demonstrate this issue using the same  settings as the base case (11 surveys, all sources). Fig.~\ref{fig:lambda_inverted} shows the inverted values for the first Lam\'e parameter $\lambda$ from survey 2 to 10. Although we observe that the general shape of $\text{CO}_2$ plume is reconstructed, artifacts due to limited aperture and strong artifacts at source locations exist. \textcolor{black}{Fig.~\ref{fig:den_sep_inverted} shows the inverted density models from survey 2 to 10, which are corrupted by strong artifacts. This phenomenon is well-known in FWI as the phase information in the data is not sensitive to density.}

We then fitted the flow equation using $\ell_2$-norm using the inverted Lam\'e $\lambda$ values as the data. \added{Since we attempted to obtain the best decoupled inversion result for comparison, we did not use the inverted density models due to the prevalent strong artifacts.} To reduce the effects of artifacts at the seismic source and receiver locations, we excluded the left most and right most parts that include the sources and receivers for inversion, and obtained the result shown in Fig.~\ref{fig:Inversion_result_sep_fit}(b). Compared to the model from coupled inversion (Fig.~\ref{fig:Inversion_result_sep_fit}(a)), the permeability model from decoupled inversion is heavily contaminated with artifacts, although we can still identify the highly permeable layer. This model provides misleading information and would be harder to interpret. Also, the inverted model has a much larger MSE (1372.24) than the base case ($\text{MSE}=218.71$). This experiment demonstrates the advantage and necessity of performing the coupled inversion.

\begin{figure}[htpb]
	\begin{center}
		\setlength{\tabcolsep}{0.2cm}
		\begin{tabular}{l}
			\small{(a)}\\
			\includegraphics[height=5cm]{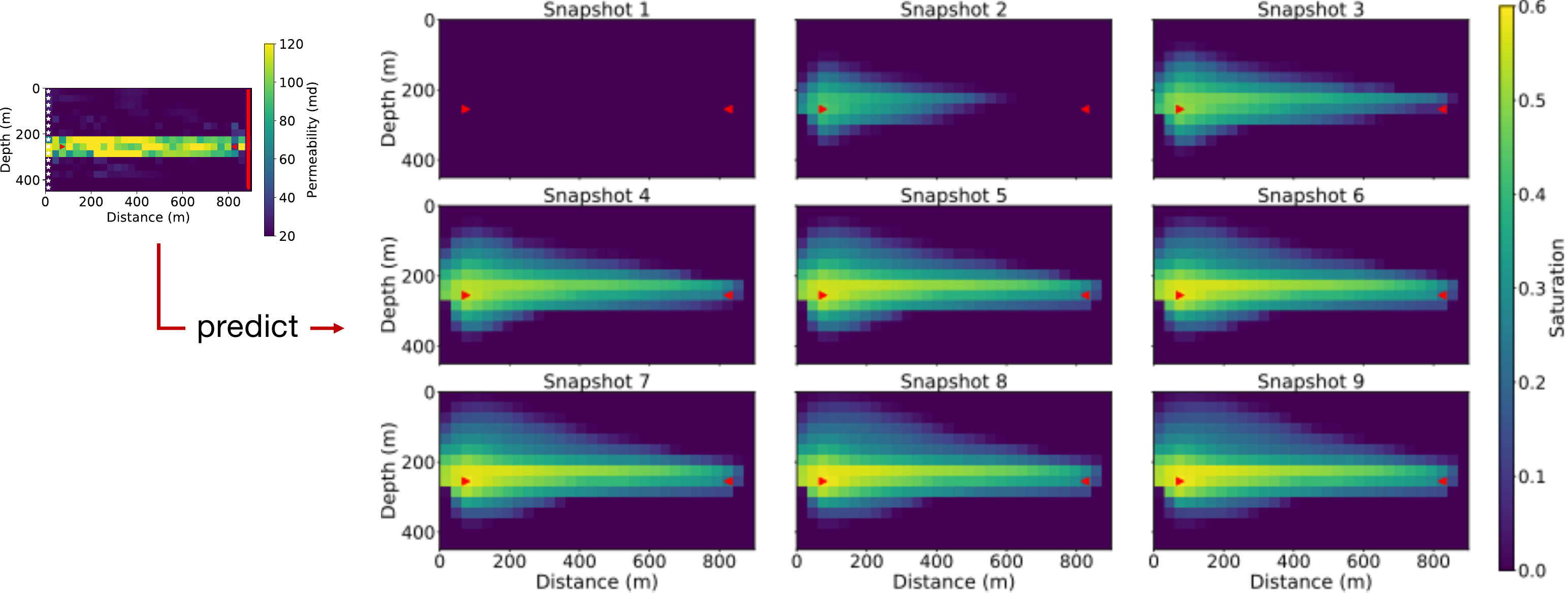}
            \end{tabular}
            \begin{tabular}{l}
			\small{(b)}\\
			\includegraphics[height=5cm]{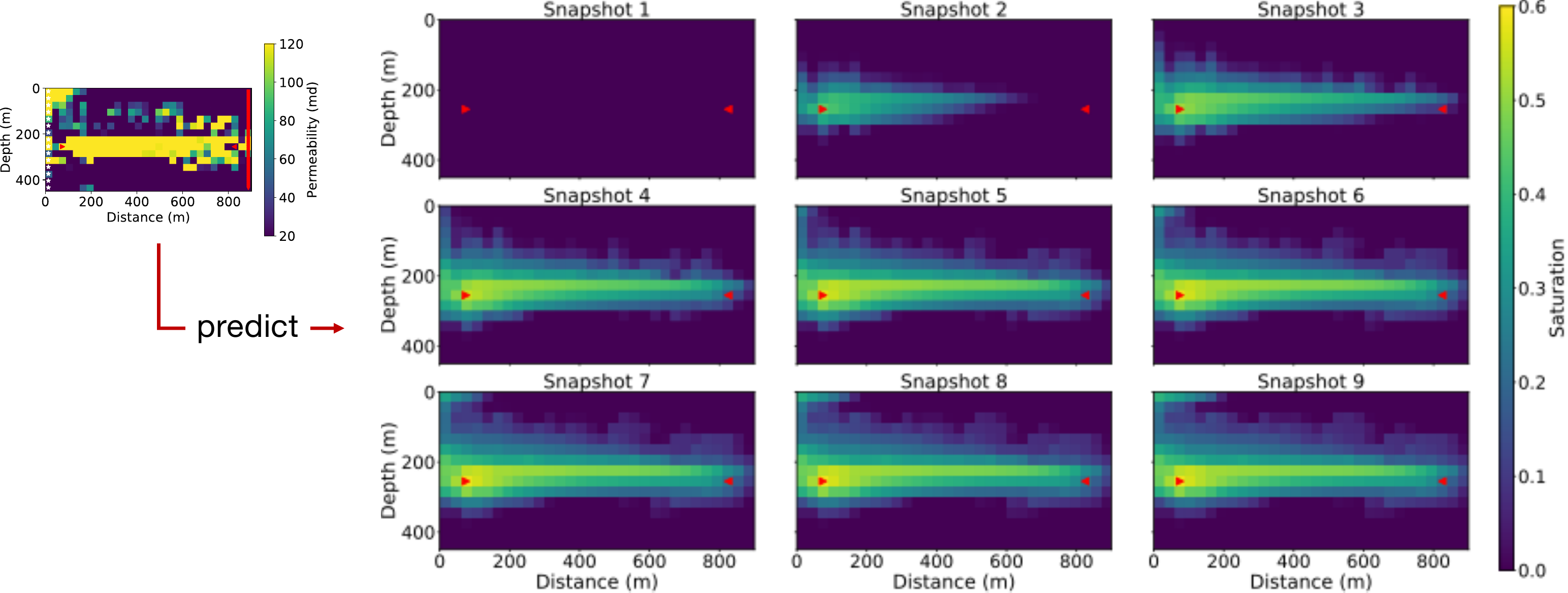}
		    \end{tabular}
		    \begin{tabular}{ll}
		    \small{(c)} & \small{(d)}\\
            \includegraphics[height=4cm]{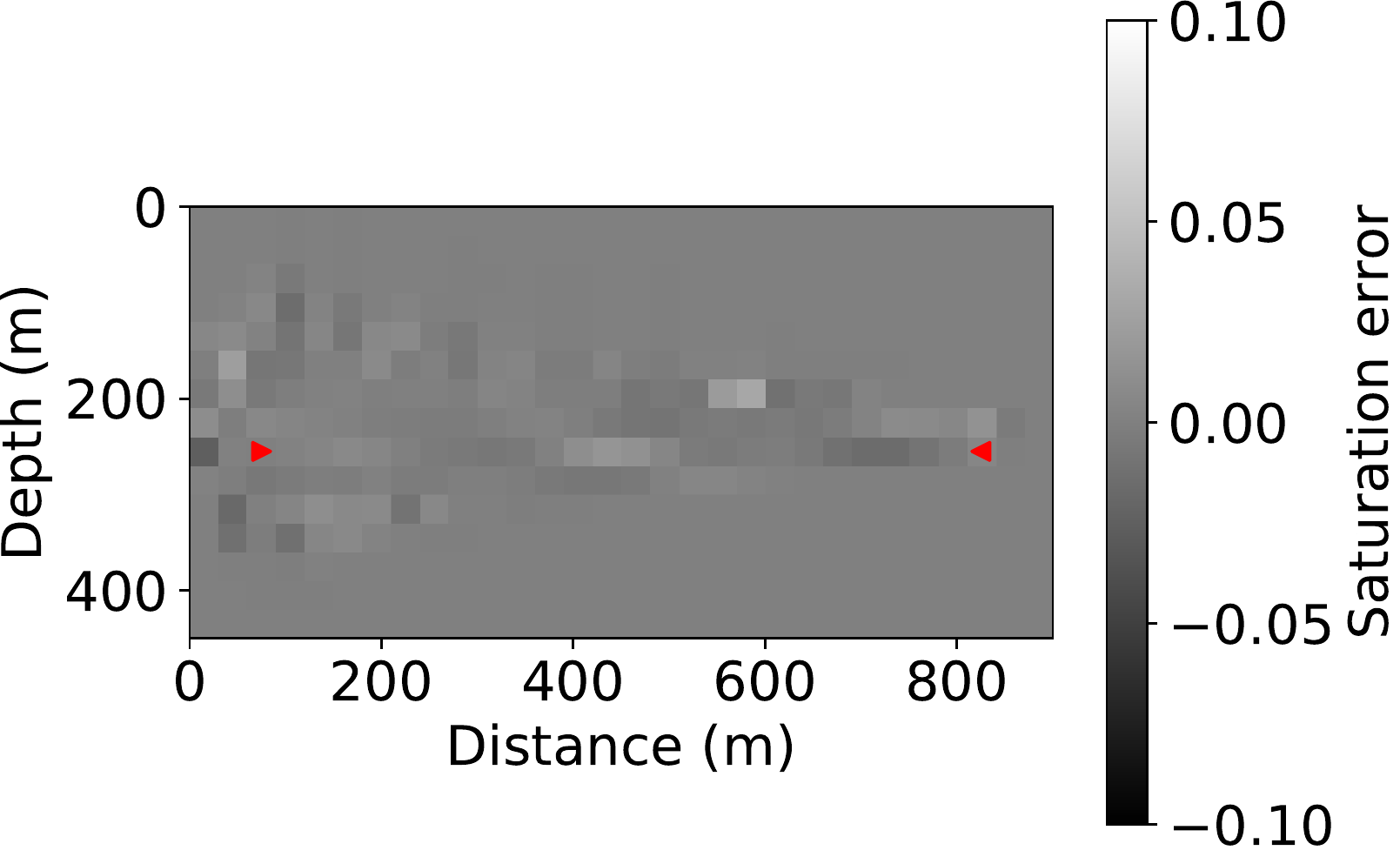} &
			\includegraphics[height=4cm]{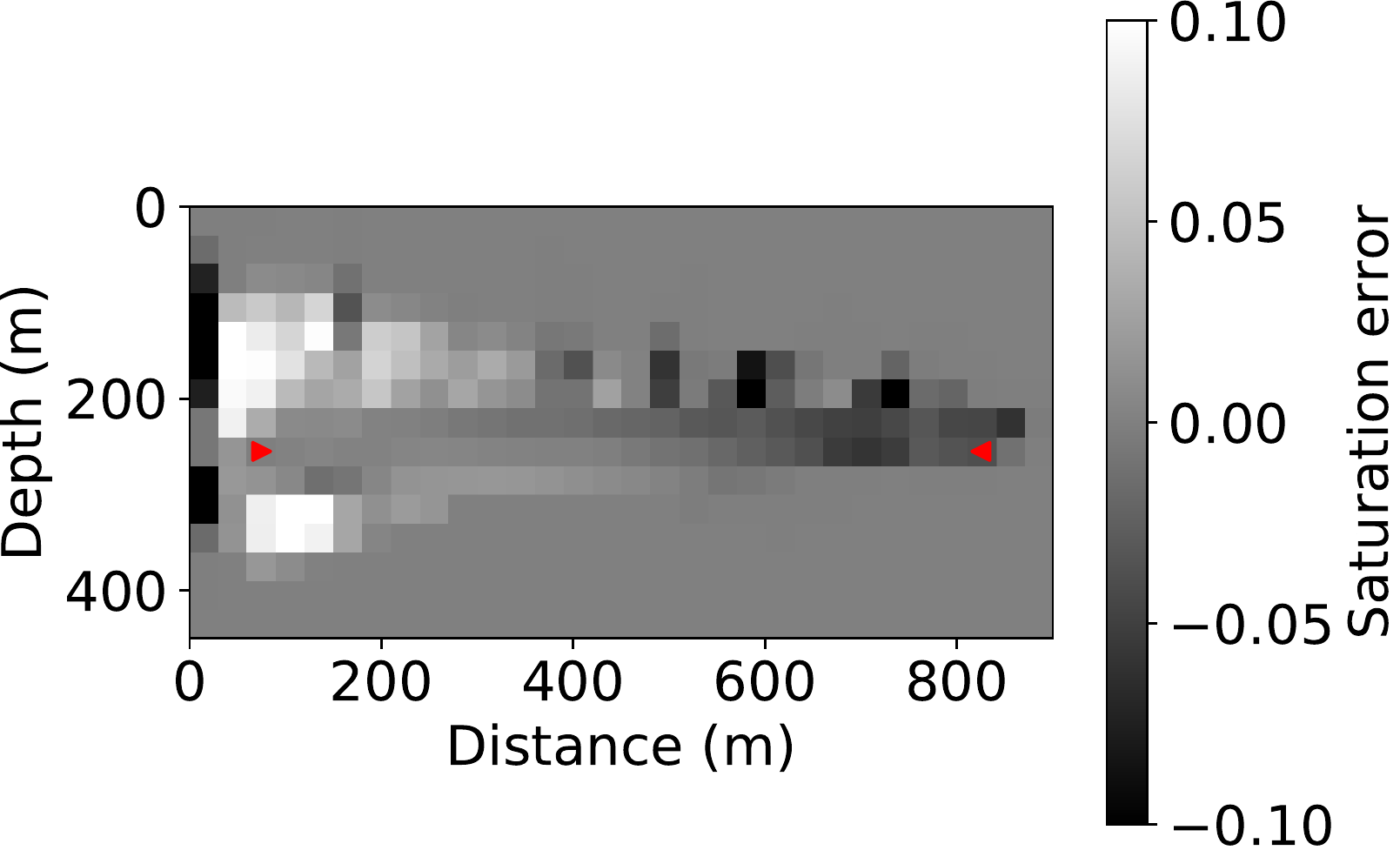}
		    \end{tabular}
	    \end{center}
	\caption{Predicted \(\text{CO}_2\) saturation with models from (a) coupled and (b) decoupled (traditional) inversion. There are strong artifacts in the predicted snapshots in (b). \added[id=r2]{We show in (c) and (d) the errors between the ground-truth saturation and the two predictions in (a) and (b) at survey 10, respectively.}}
	\label{fig:predicted_sat_evo}
\end{figure}

The inverted intrinsic properties can be used to predict flow behavior and also help reservoir management and optimization. With the inverted models from both the coupled and decoupled inversion strategy, we perform forward simulations of the flow and convert saturation maps to P-wave velocity maps. Fig~\ref{fig:predicted_sat_evo}(a) shows the predicted P-wave velocity evolution in slow time with the coupled inversion result, while Fig~\ref{fig:predicted_sat_evo}(b) shows the snapshots with the decoupled inversion result. The former one \added[id=r3]{yields} better predictions that are closer to the ground truth (Fig.~\ref{fig:Sat_evo_patchy_true}) while the latter one leads to erroneous predictions with strong artifacts. \added[id=r2]{Fig.~\ref{fig:predicted_sat_evo}(c) and (d) show that the error between the true saturation and the prediction from the decoupled inversion result is much larger than that of the coupled inversion. To further illustrate our point, we conducted forward simulations with the predicted elastic models, and compute the sum of the misfit between the predicted data and the observed data at each snapshot. The data misfit is 0.3097 for the coupled inversion, while it is 181.6226 for the decoupled inversion, which indicates that our coupled inversion can fit the data much better than the decoupled strategy.} 

{\color{black}
\subsection{FWI and Eikonal Equation Inversion}
Full waveforms contain much more information than mere traveltime. As a result, FWI can produce results of much higher resolution than traveltime-based tomography methods such as the ray tomography. In this section, we compare the effects of using waveform data with first-arrival time only on the coupled inversion. Instead of using the traditional ray-tomography, we cast the first-arrival traveltime inversion as a PDE-constrained inversion as well using the Eikonal equation~\cite{aki2002quantitative}, so that we can apply our intrusive automatic differentiation (IAD) technique to this inversion problem. This formulation also reduces the error from ray-tracing, allowing a better comparison with FWI. In the synthetic tests, we assume that the Eikonal equation can accurately predict the first-arrival time. However, we should keep it in mind that the conventional ray-tomography and the Eikonal tomography are based upon the high-frequency approximation of wave propagation~\cite{yilmaz2001seismic}; thus, there are additional errors from this inaccurate assumption in physical models. Also, there may be errors from first-arrival picking when dealing with field data. 

The Eikonal equation is given by 
\begin{align}\label{equ:Eikonal}
& |\nabla T(\bx;\mathbf{x}_s)| = \frac{1}{v(\bx)}, \quad \bx \in \RR^2, \\ \nonumber
& T(\mathbf{x}_s;\mathbf{x}_s) = 0,
\end{align}
where $T(\bx;\mathbf{x}_s)$ is the first arrival time at location $\bx$ for a wave front starting at the source location $\mathbf{x}_s$, computed with the velocity field $v(\bx)$. We solve the Eikonal equation~(\ref{equ:Eikonal}) using the fast sweeping method~\cite{zhao2005fast}, which solves a system of discrete nonlinear equations at interior grid points by applying a sequence of Gauss-Seidel iterations:
$$[(T_{i,j}^h-T_{x, \min}^h)^+]^2 + [(T^h_{i,j} - T^h_{y, \min})^+]^2 = \frac{1}{v_{i,j}^2} h^2$$
where $T_{x, \min}^h = \min(T_{i-1,j}^h, T_{i+1,j}^h)$, $T_{y, \min}^h = \min(T_{i,j-1}^h, T_{i,j+1}^h)$, $x^+ = \max(0, x)$, and $h$ is the grid spacing.

We conducted the inversion using the IAD technique, where the numerical solver and its automatic differentiation rule are implemented using a custom operator. Fig.~\ref{fig:fwi_eikonal} shows a comparison of the coupled inversion using the Eikonal equation and FWI. We see that the Eikonal result has stronger artifacts in the white circled area, although the two results are comparable. \Cref{tab:fwi_eikonal} summarizes a further comparison between these two methods. The result associated with FWI is much more accurate, especially when the number of sources is small, suggesting that the information beyond the first-arrival time in seismic waveforms is helpful to calibrate permeability more accurately. This is important for monitoring projects using sparsely distributed sources and receivers.
\begin{figure}[htpb]
      \begin{center}
            \setlength{\tabcolsep}{0.2cm}
            \begin{tabular}{l l}
                  \small{(a)} & \small{(b)}\\
                  \includegraphics[height=4cm]{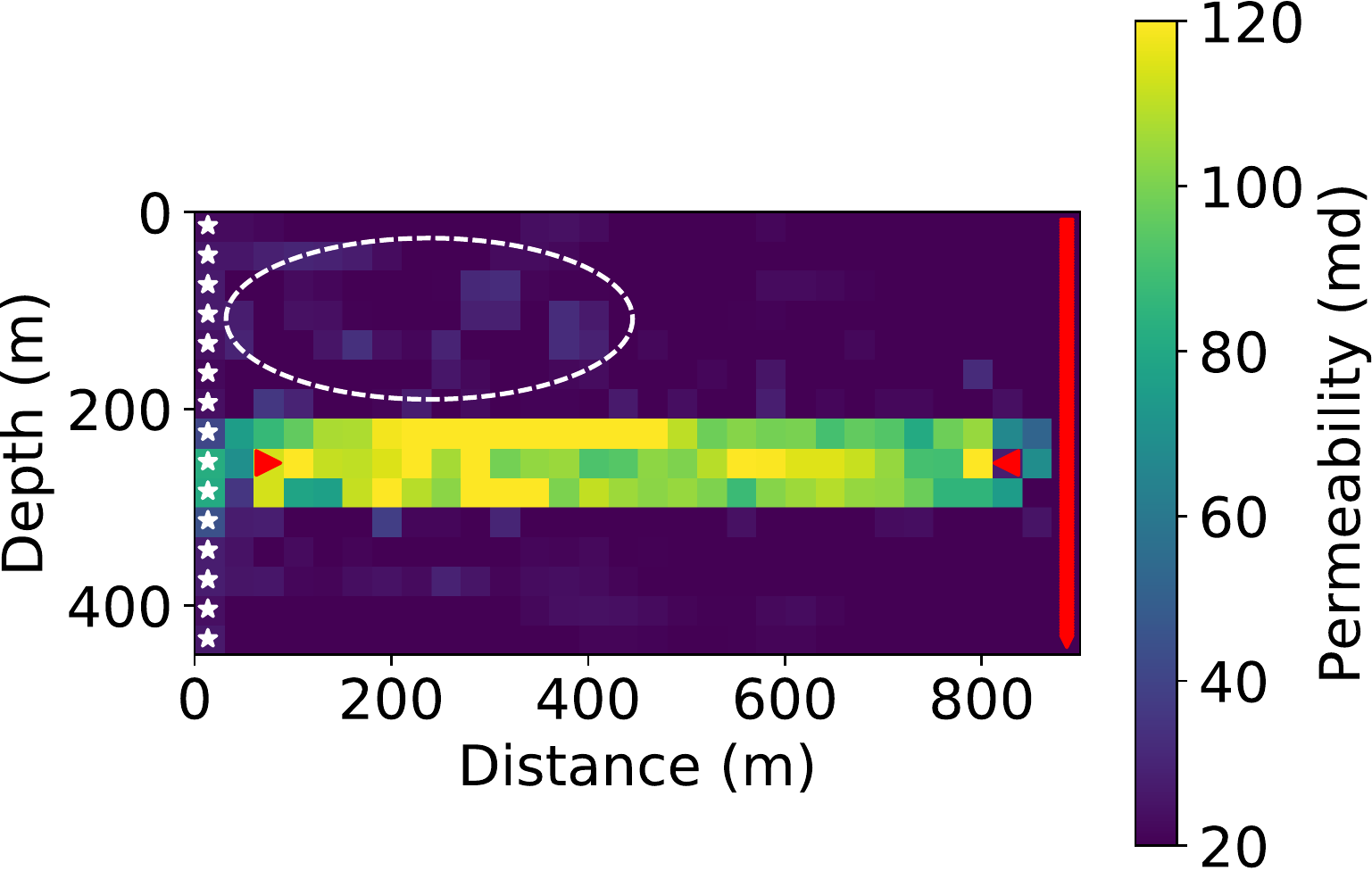} & \includegraphics[height=4cm]{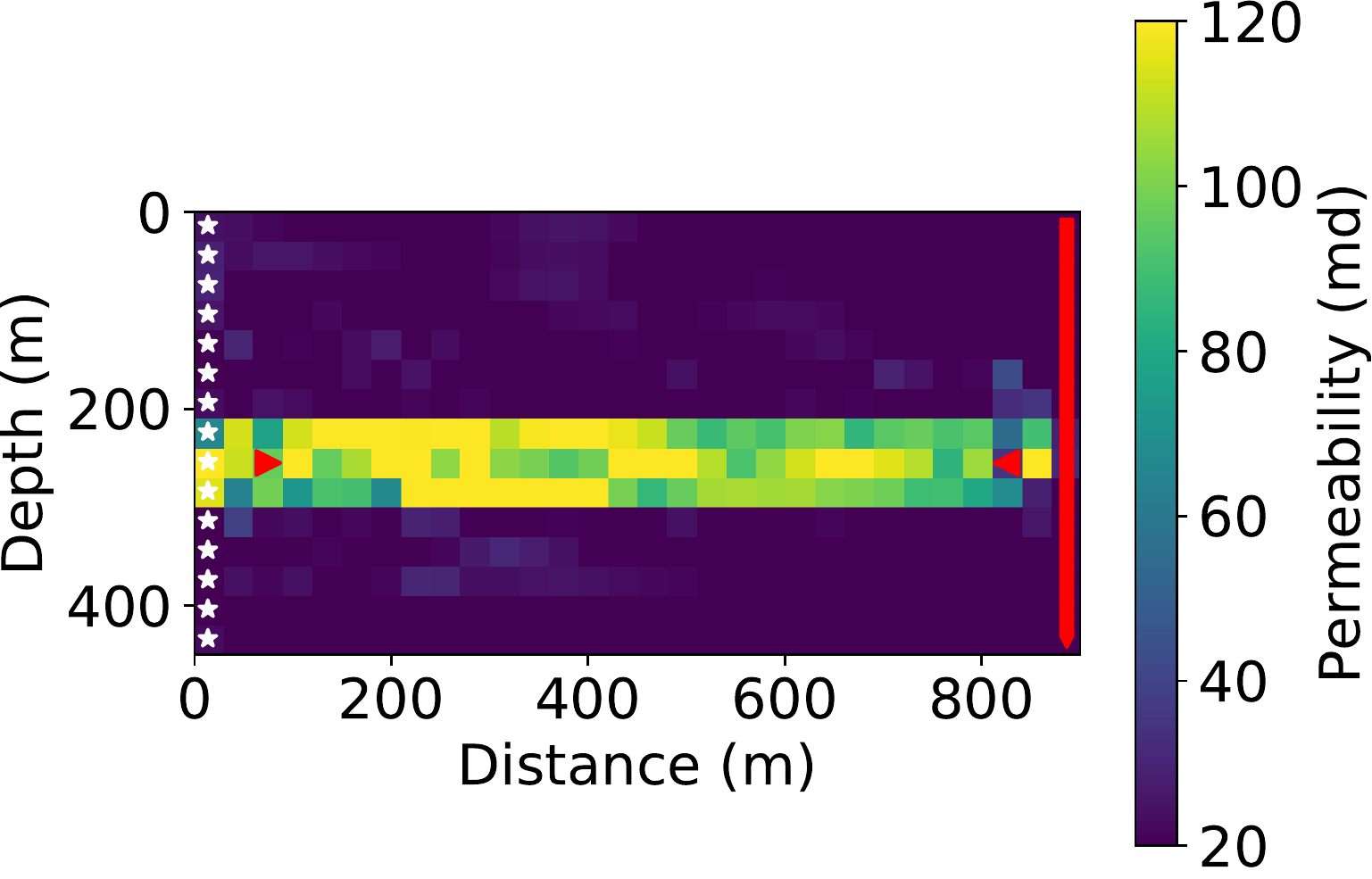} \\
            \end{tabular}
      \end{center}
      \caption{Permeability estimated from coupled inversion with the Eikonal equation (left) and FWI (right). We see that the result associated with the Eikonal equation has more artifacts (e.g., in the white circled area), indicating that the richer information in full waveforms can help invert permeability more accurately. }
      \label{fig:fwi_eikonal}
\end{figure}

\begin{table}[htpb]
      \caption{MSE of inverted permeability models from FWI and the Eikonal equation inversion. The result associated with FWI is much more accurate, especially when the number of sources is small.}
      \centering
      \begin{tabular}{c c c c}
      \hline
       & \textbf{15 sources} & \textbf{3 sources} & \textbf{1 source}   \\
       \textbf{FWI} & 218.71 & 245.94 & 294.60 \\
       \textbf{Eikonal} & 271.59 & 415.74 & 479.78 \\
      \hline
      \end{tabular}
      \label{tab:fwi_eikonal}
      \end{table}
}

\subsection{Inversion of Rock Physics Parameters}\label{subsect:brie_inversion}
\begin{figure}[htpb]
\centering
\includegraphics[width=0.6\textwidth]{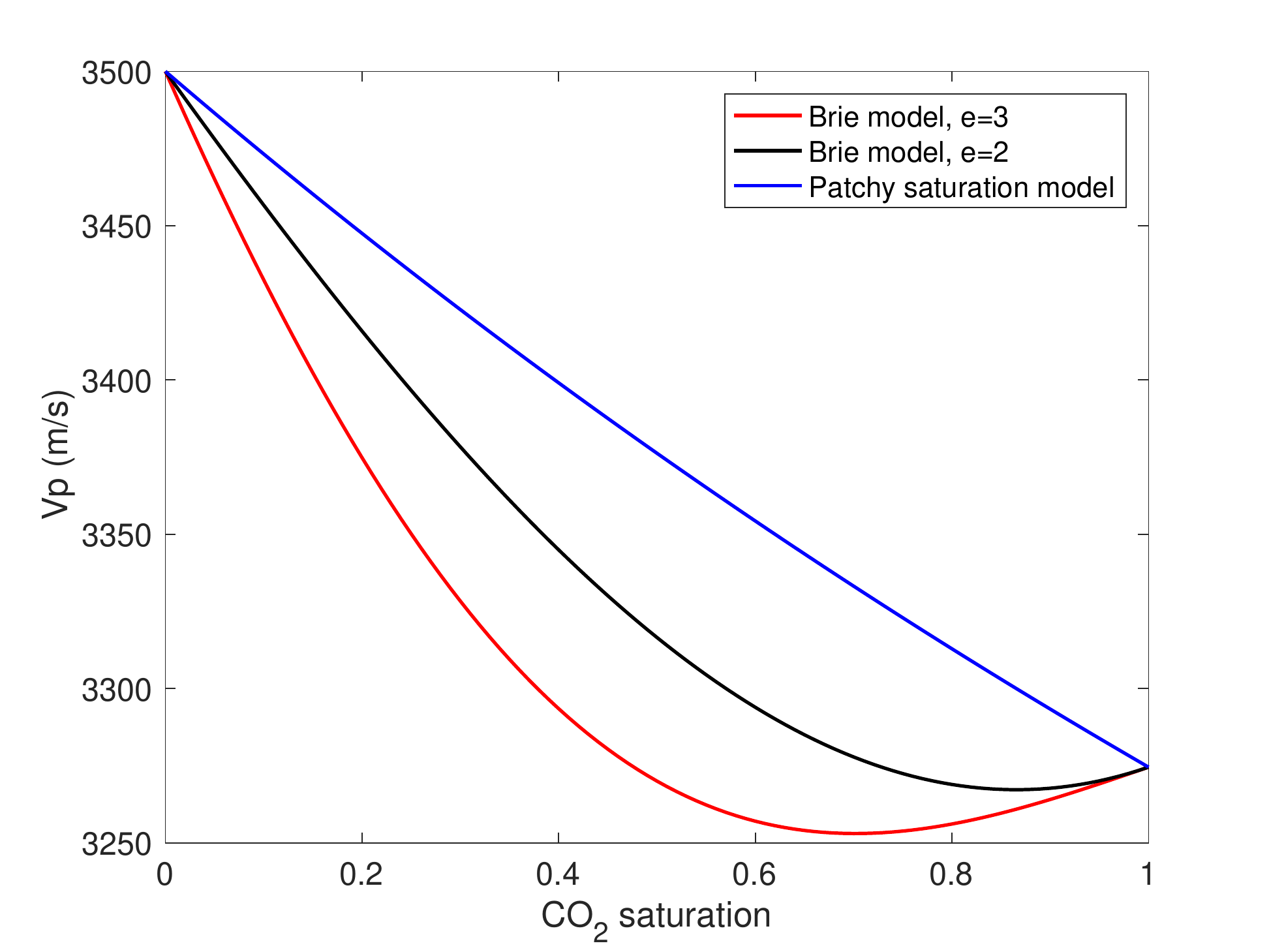}
\caption{The $V_p$-saturation curve of the patchy saturation model, and the Gassmann's model with Brie's fluid mixture equations with different coefficients ($e=2$ and $e=3$).}
\label{fig:rock_physics_curves}
\end{figure}

The rock physics model is a type of closure relationship that is sometimes empirical. We want to examine whether our method can also provide insight on calibration or discovery of such relationships. Therefore, we switched the rock physics model from the patchy saturation model to Gassmann's model with Brie's fluid mixture equation. The patchy saturation model is the upper bound of responses of elastic properties to fluid substitution. The Brie model introduced in SI 1.2.2, on the other hand, can approximate a wide range of rock physics models parameterized with the coefficient $e$. We plot the P-wave velocity of rocks against the saturation of $\text{CO}_2$ with the same parameters in previous numerical experiments. As is shown in Fig.~(\ref{fig:rock_physics_curves}), $V_p$ drops first and then increases as the saturation of $\text{CO}_2$ increases in the Brie model, and as the coefficient $e$ decreases, the curve gradually approximates the upper bound, a straight-line predicted by the patchy saturation model. 

\begin{figure}[htpb]
	\begin{center}
		\setlength{\tabcolsep}{0.2cm}
		\begin{tabular}{l l}
			\small{(a)} & \small{(b)}\\
			\includegraphics[height=5cm]{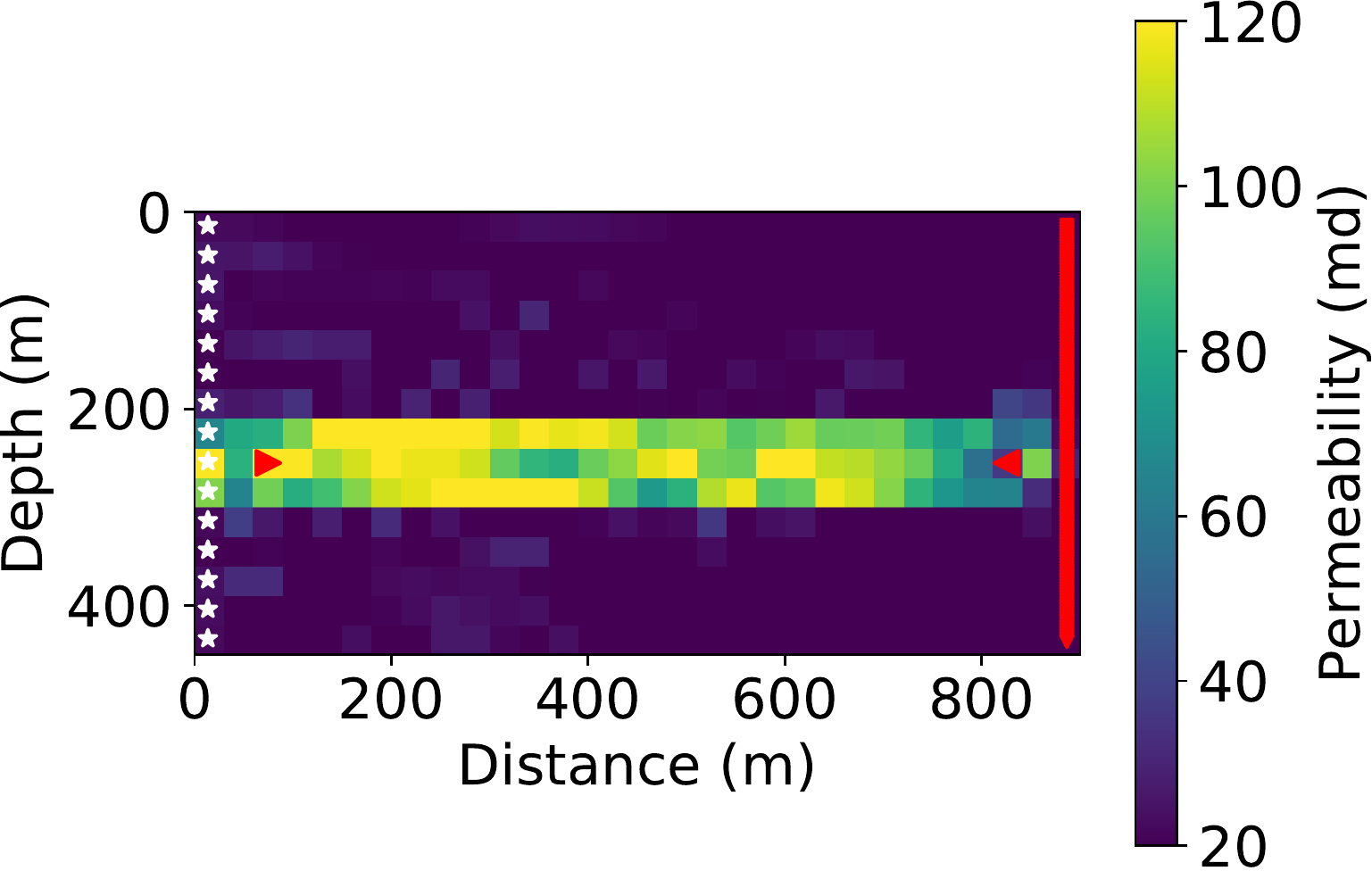}
			&  \includegraphics[height=5cm]{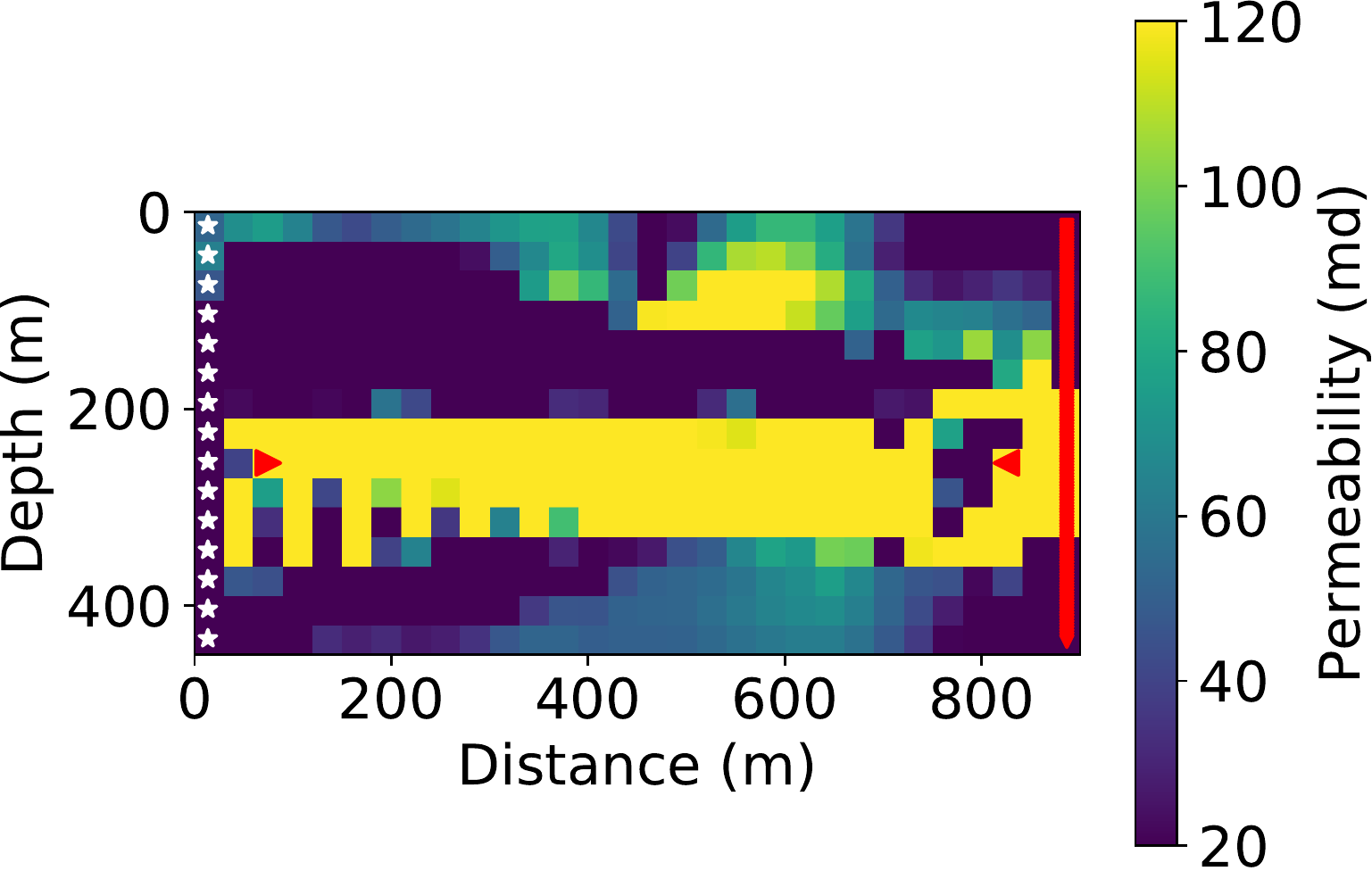}
		\end{tabular}
		\begin{tabular}{l l}
			\small{(c)} & \small{(d)}\\
			\includegraphics[height=5cm]{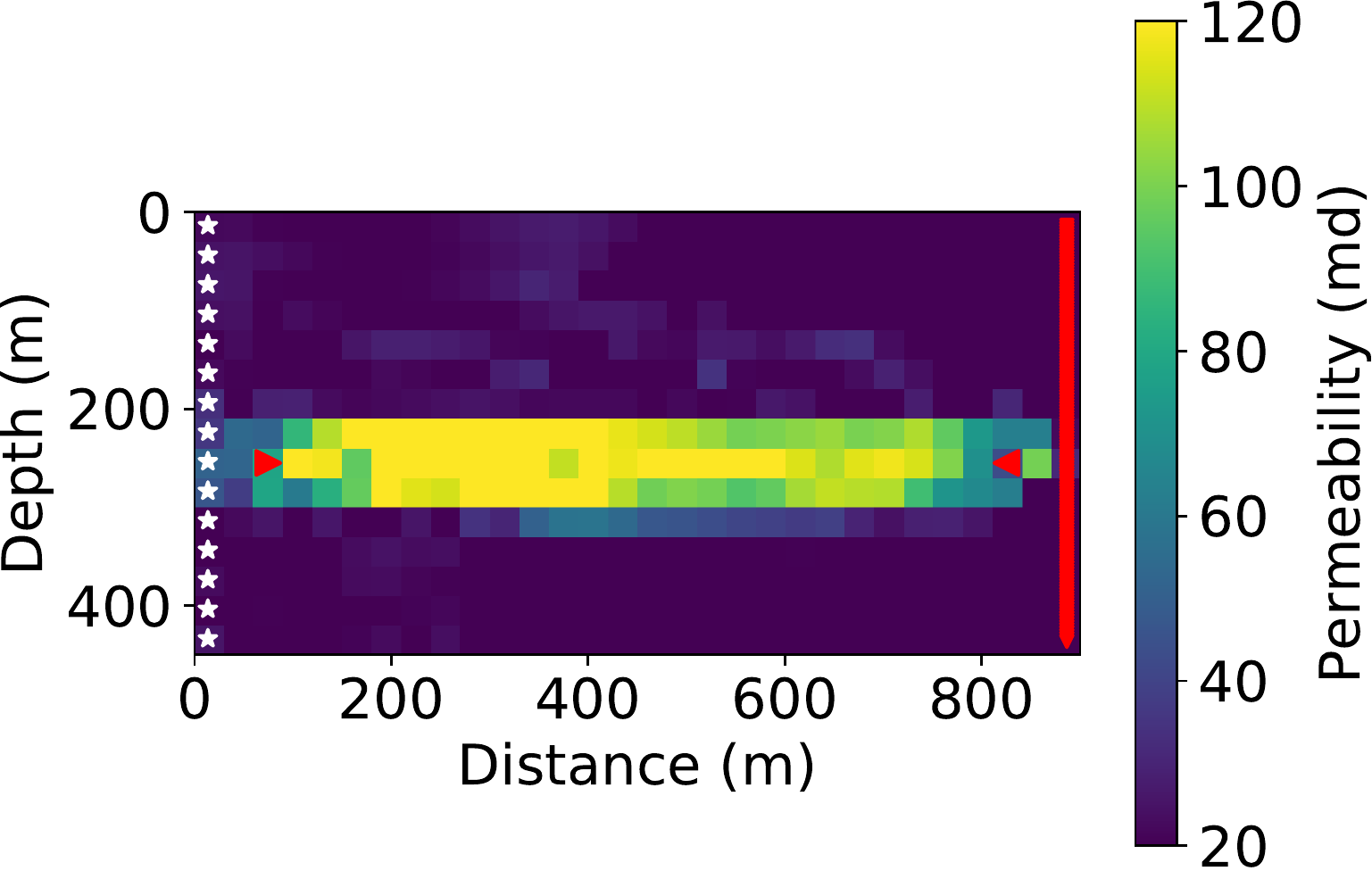}
			& \includegraphics[height=4.5cm]{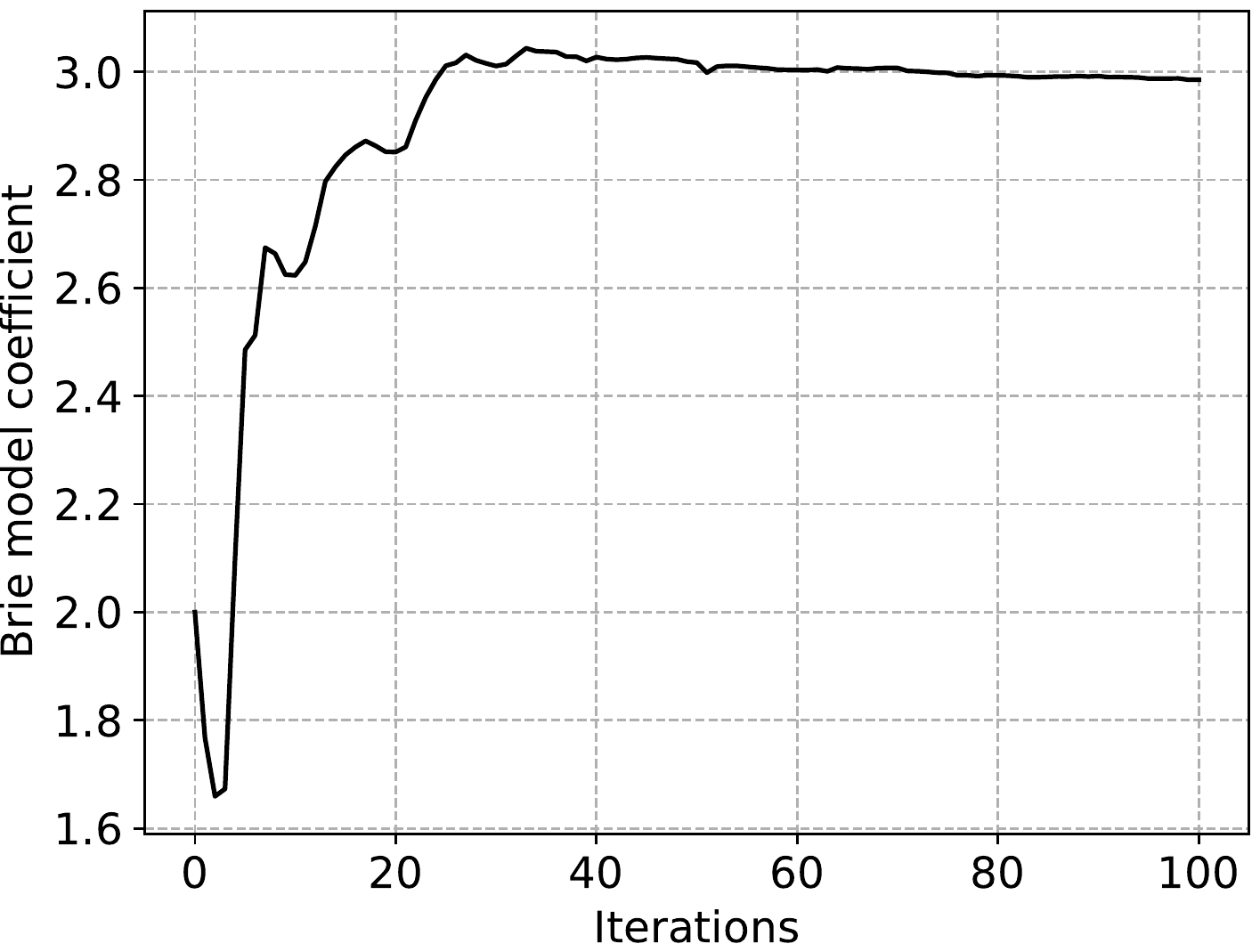} 
            \end{tabular}\\
            \begin{tabular}{l}
                  \small{(e)}\\
                  \includegraphics[height=5cm]{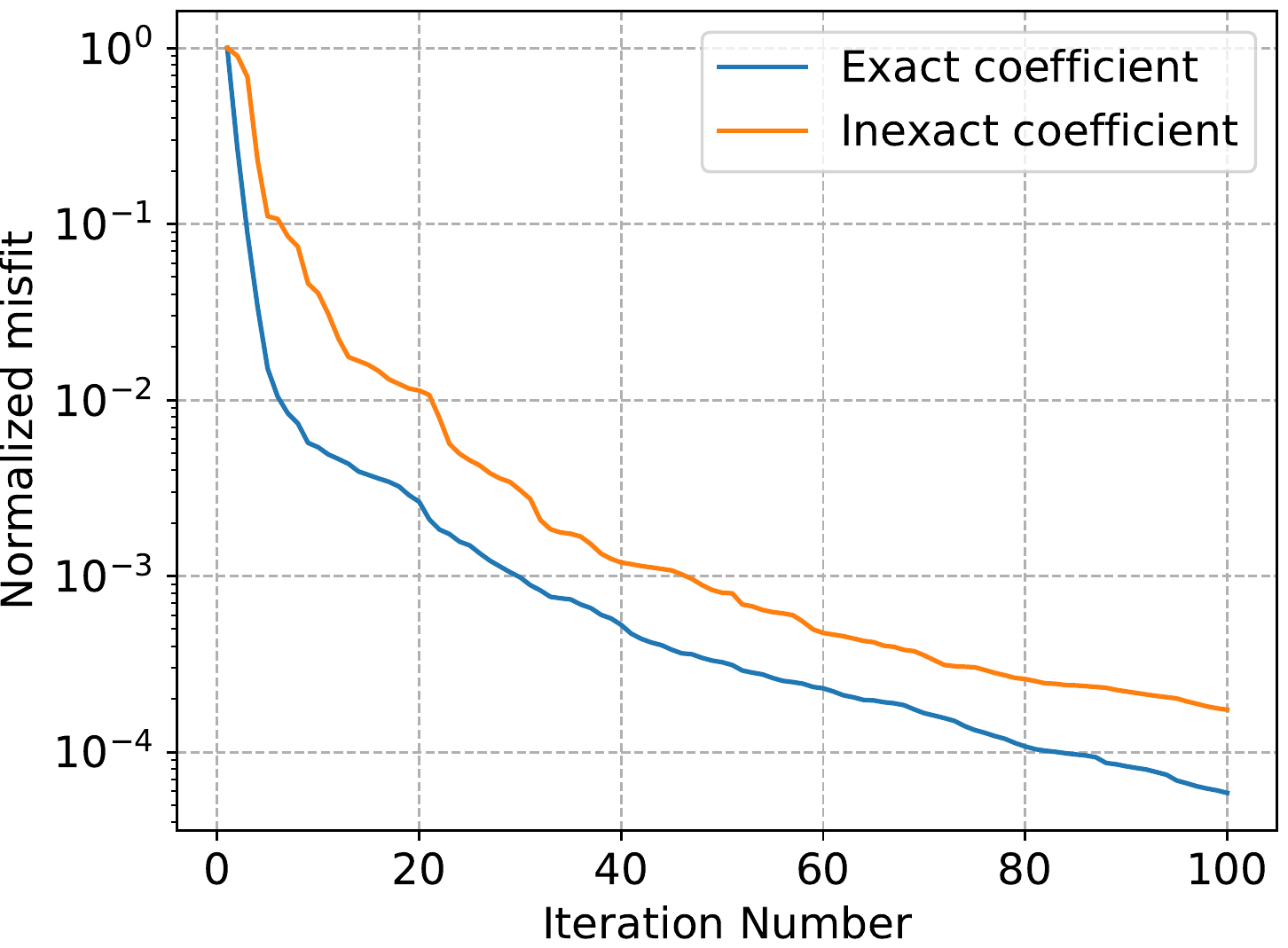}
            \end{tabular}
	\end{center}
	\caption{The inversion results with the Brie model. (a) Inverted permeability with the exact coefficient ($e$=3); (b) Inverted permeability with incorrect coefficient ($e$=2); (c) Inverted permeability with incorrect initial coefficient ($e$=2) which is updated simultaneously; (d) The evolution of the value of $e$ with iterations; (e) The convergence curves of the objective function for both cases.}
	\label{fig:Inversion_result_brie}
\end{figure}

In the inversion experiments, the true rock physics model was assumed to be the Brie model with $e=3$. We first used the exact rock physics model and Fig.~\ref{fig:Inversion_result_brie}(a) shows the inversion result, which has comparable accuracy as that from the patchy saturation model, with a low MSE equals 250.64. This result relies on the accurate assumption of rock physics models. However, if we have little information about the true rock physics model and wrongly assume that $e=2$, Fig.~\ref{fig:Inversion_result_brie}(b) shows the inverted result that has much stronger artifacts and much higher MSE (2098.02). To mitigate such a common problem, we invert for the Brie's coefficient \(e\) and permeability simultaneously. Thanks to the automatic differentiation capability of the framework, the gradient of the misfit function with respect to \(e\) can be easily obtained. The initial guess of \(e\) was also 2 as the previous experiment. Considering the range of value of the \(e\) coefficient is different from that of the permeability, we scaled up the value of \(e\) parameter by 30 and divided it by 30 before feeding it to the rock physics equations. Fig.~\ref{fig:Inversion_result_brie}(c) shows the inverted permeability with \(e\) updated simultaneously, which is more accurate (MSE = 320.04) than the previous case without updating. We show the inversion history of the \(e\) coefficient in Fig.~\ref{fig:Inversion_result_brie}(d), which quickly converges to the true \(e=3\) after around 40 iterations. The initial drop of the value at the beginning several iterations are probably caused by the fact that the L-BFGS-B optimizer essentially acts as a gradient-descent at the beginning, which cannot handle the scaling of different parameters of different physical meanings well. This phenomenon was severe without the scaling trick that we just mentioned. In Fig.~\ref{fig:Inversion_result_brie}(e), we observe good convergence in data misfit of the simultaneous inversion. This promising result not only presents the multi-parameter inversion capability enabled by the automatic differentiable computing but also emphasizes that the simultaneous inversion can yield more accurate inversion results and provide insights on empirical relationships in practice.

\subsection{A 2-D Stream Channel Model}\label{subsect:progressive_inversion}
\begin{figure}[htpb]
	\begin{center}
		\setlength{\tabcolsep}{0.2cm}
		\begin{tabular}{l l}
			\small{(a)} & \small{(b)}\\
			\includegraphics[width=0.48\textwidth]{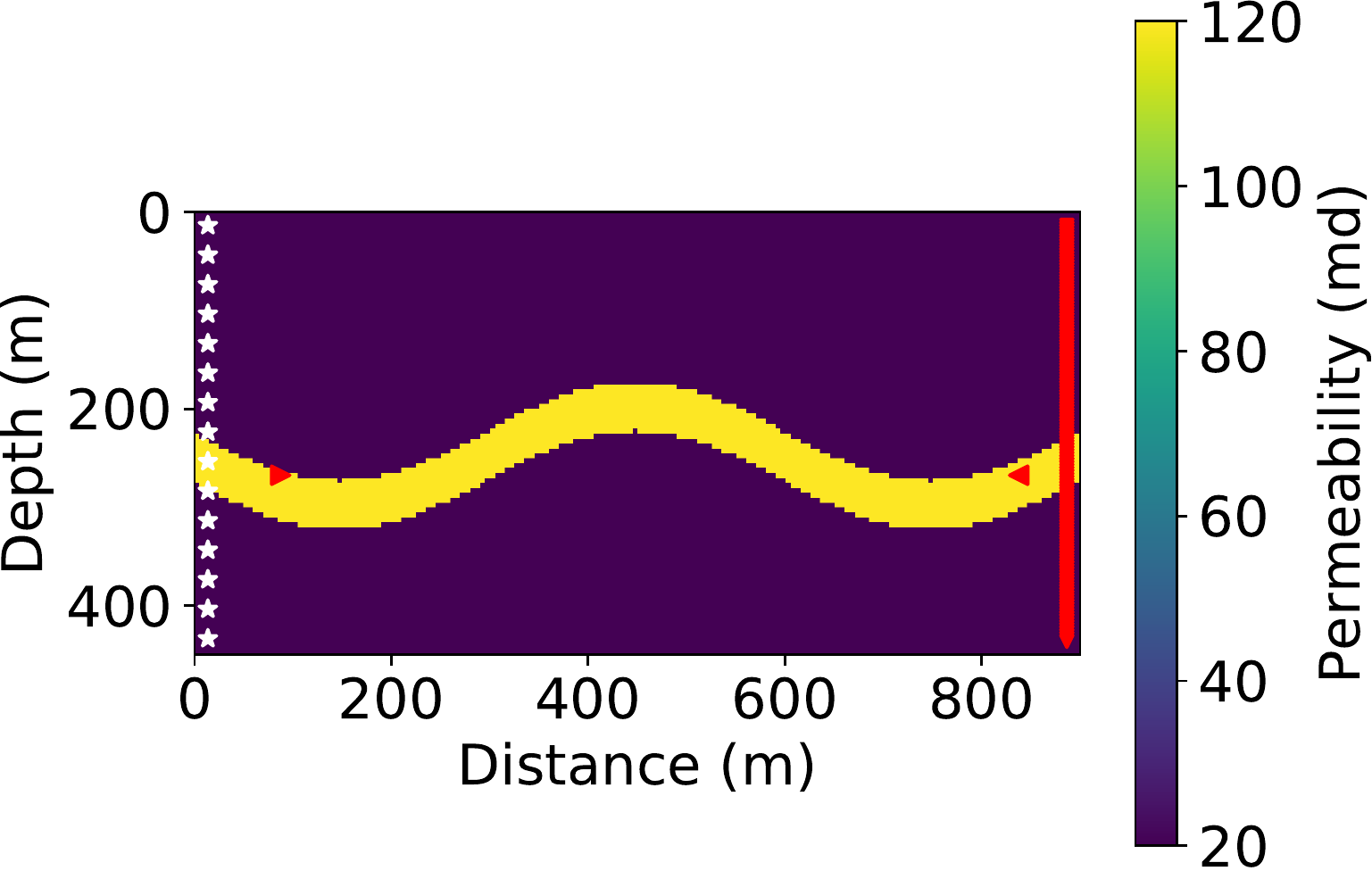}
			&  \includegraphics[width=0.48\textwidth]{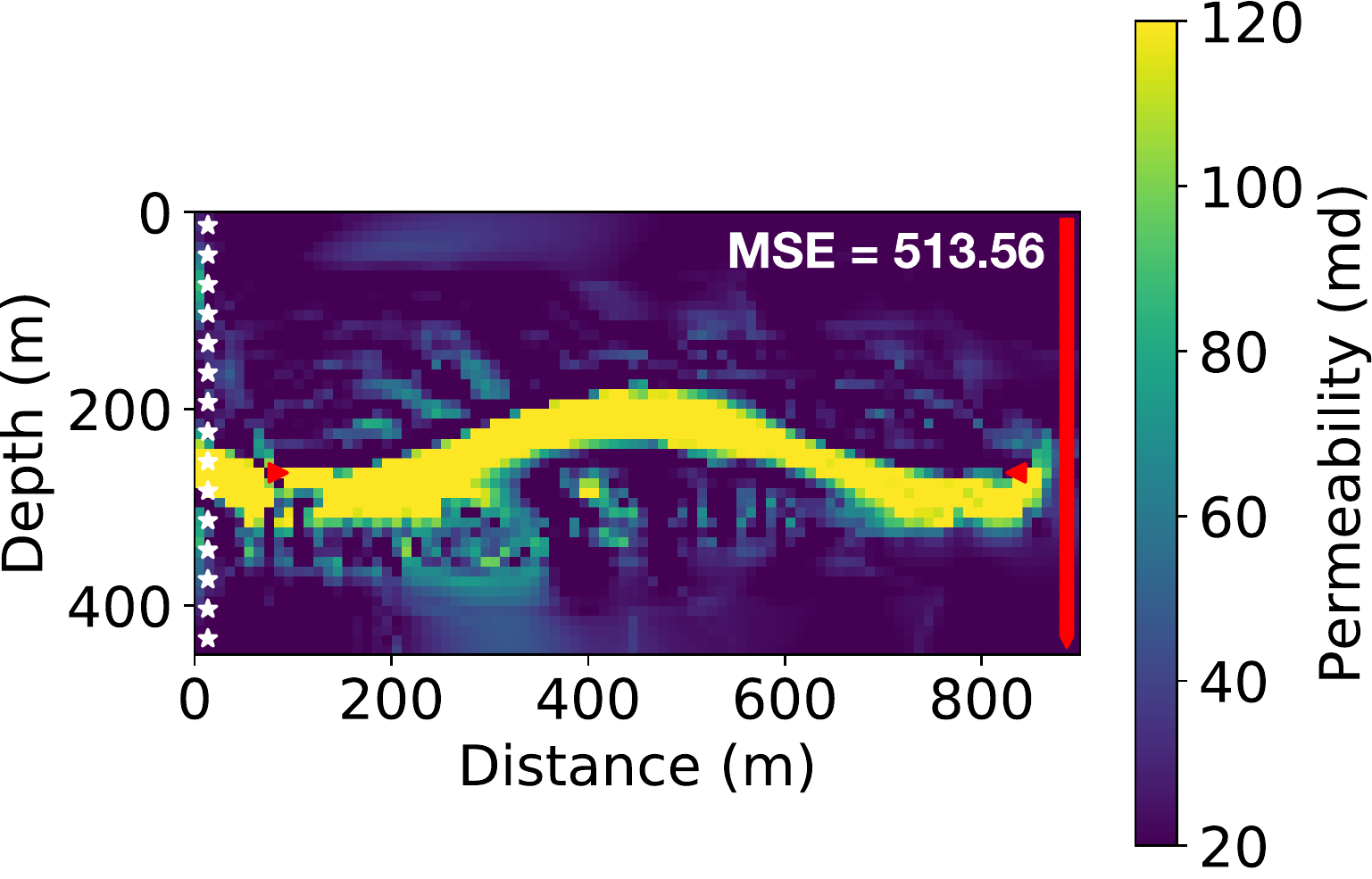}
            \end{tabular}
		\begin{tabular}{l}
			\small{(c)}\\
			\includegraphics[width=0.48\textwidth]{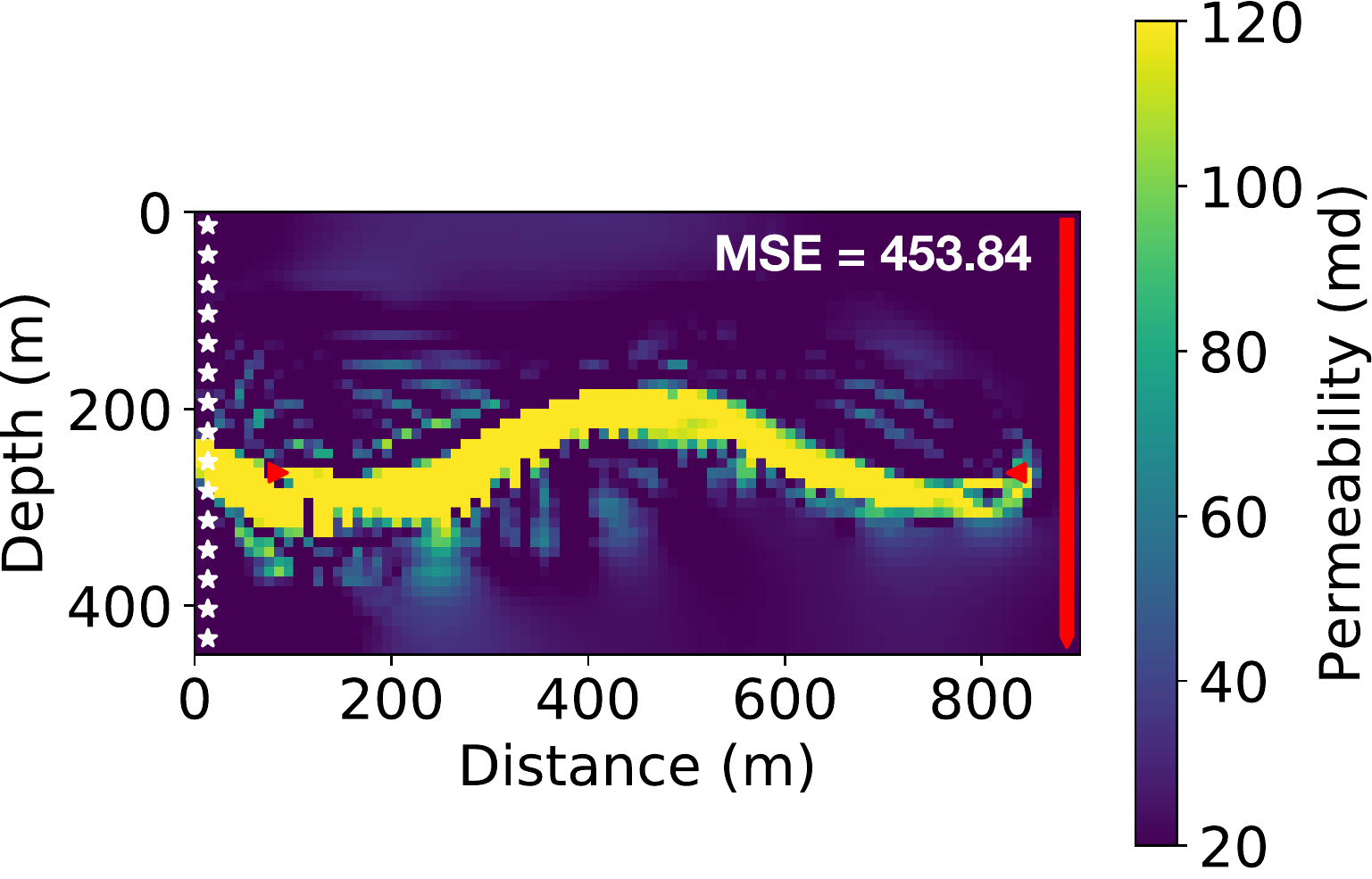}
		\end{tabular}
	\end{center}
	\caption{The inversion results of a channel model. (a) The true permeability model; (b) the inverted permeability model with the all-at-once inversion strategy; (c) the inverted permeability model with the continuous inversion strategy. The initial model is homogeneous with 20 md permeability as previous experiments.}
	\label{fig:channel_inv_pgs}
\end{figure}

In addition to the 3-layer model of horizontal layers, we also tested the coupled inversion for a stream channel model shown in Fig.~\ref{fig:channel_inv_pgs}(a), where the channel has a permeability of 120 md while the background is 20 md. This example can represent a vertical slice between vertical wells or a horizontal slice between horizontal wells.

\begin{figure}[htpb]
      \noindent\includegraphics[width=\textwidth]{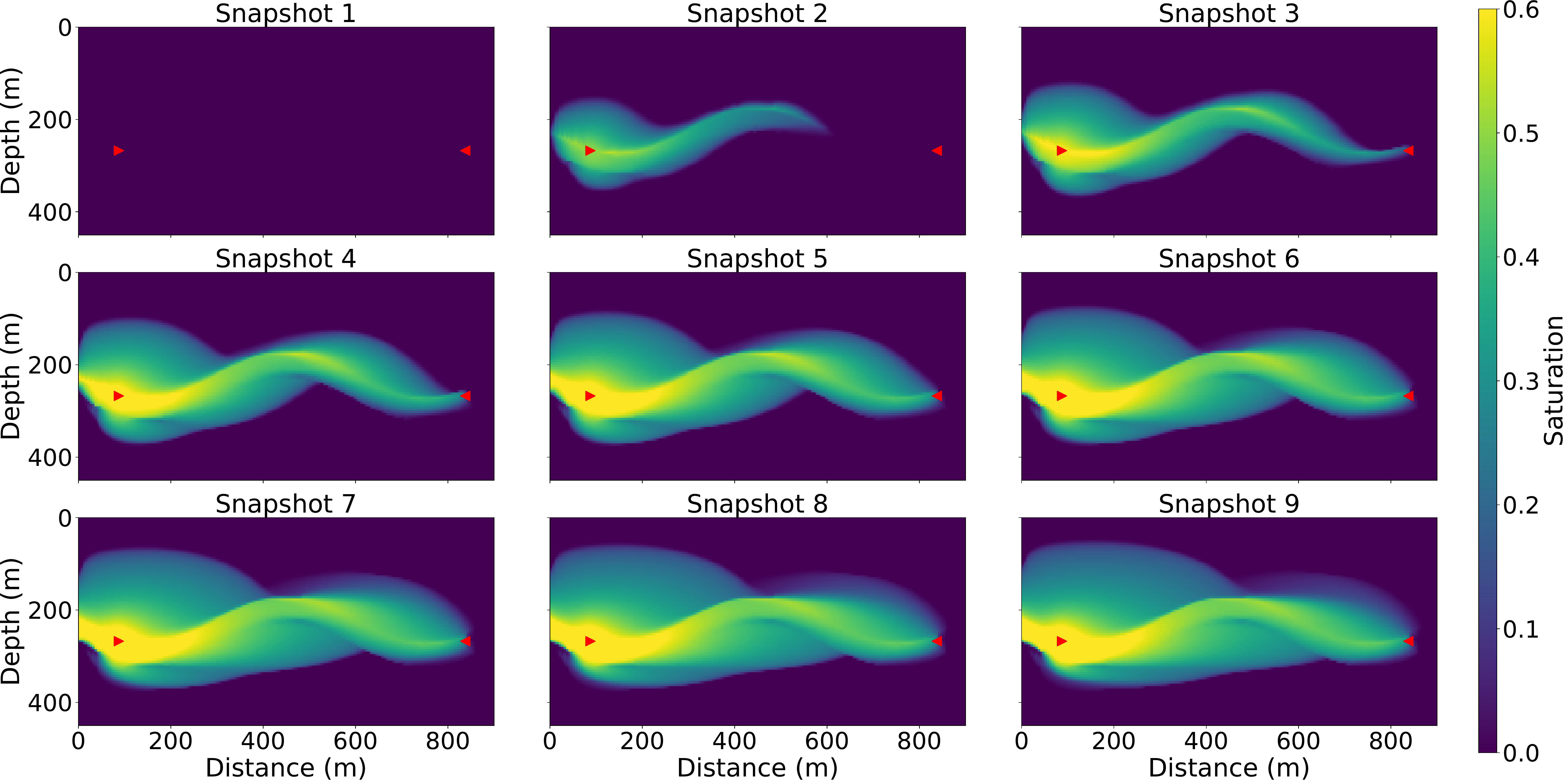}
      \caption{The first 9 snapshots of $\text{CO}_2$ saturation evolution of the channel model.}
      \label{fig:sat_evo_patchy_true_channel}
\end{figure}

This model has the same dimension as the 3-layer model, but has a smaller grid size of 5 m in both the vertical and horizontal directions, hence the model is parameterized by \(90 \times 180\) cells. We used this discretized model to create synthetic observed data. The synthetic saturation evolution maps for this channel model is shown in Fig.~\ref{fig:sat_evo_patchy_true_channel} for nine snapshots in slow-time. For the inversion, we increase the grid size to 10 m yielding a grid of \(45 \times 90\) cells. The initial model is a homogeneous model of permeability of 20 md as previous experiments. Fig.~\ref{fig:channel_inv_pgs}(b) show the inversion result with data from 11 surveys after 100 iterations, which is the same all-at-once strategy as previous experiments. The 2-D channel is well reconstructed, meaning that our algorithm can also resolve complex 2-D structures with denser parameterization. The artifacts seen can be reduced by geological parameterization~\cite{landa1997procedure}, or regularization methods~\cite{Aster201393,li2018full}. 

In real-world operations, it would be rare to wait and begin data analysis after completion of all surveys. Instead, it is a more practical strategy to perform continuous inversion along with new data from continuing surveys. In other words, at the end of survey \(i\, (i>=2)\), we formulate an optimization/inversion problem with all available data so far \(\mathbf{d}^{\text{obs}}_j \, (j=1,\cdots,i)\) and use the model from the previous stage of inversion as the initial model. We also tested the new strategy in this case. After a continuous inversion of only 6 surveys, we did not observe further improvement on the inversion result and report the inverted permeability model in Fig.~\ref{fig:channel_inv_pgs}(c), which seems to have fewer artifacts and lower MSE than that from the all-at-once strategy. Instead of inverting for all parameters at the same time, the continuous inversion strategy breaks the original problem into progressive bits, and it is easier to reconstruct the permeability field incrementally. Also, an accurate estimation of the flow properties at an early stage largely determines the accuracy of the following inversions due to the \replaced[id=r3]{underlying}{underline} physics model.

\added[id=r1]{
\begin{remark}
In a more realistic setting, the permeability within the channel may also be spatially varying instead of having a constant value. As an additional example, we summarize the inversion results for this scenario in the supporting information (SI 7). We show that our method is also effective for calibrating space-varying permeability within the channels. 
\end{remark}}


\section{Discussion \& Conclusions}\label{sect:discussion_conlusinons}
We have presented a novel approach for estimating intrinsic rock parameters \replaced{(e.g., permeability)}{such as permeability} from time-lapse full-waveform inversion of geophysical data \replaced{(e.g., seismic data)}{such as seismic}. We demonstrated that solving an inverse problem constrained by coupled physical systems is superior to a decoupled inversion in terms of accuracy. \added{Additionally, we found that the full waveform data carry richer information than first-arrival travel time data for calibrating hydrological properties.} Also, the method is effective in inverting an empirical rock physics model while estimating the subsurface permeability map at the same time. 

\replaced{To solve the coupled inverse problem, we proposed the intrusive automatic differentiation (IAD) method}{We presented an intelligent automatic differentiation strategy for inversion}. This strategy combines different levels of control over gradient computation schemes \added{in an intrusive way to an automatic differentiation computational workflow}: \replaced{custom}{customized} PDE operators for time step marching in solving the flow equations, built-in differentiable operators from \texttt{TensorFlow} for the rock physics models, and the discrete adjoint method for wave physics. With this strategy, we can fully exploit the power of modern machine learning platforms, and also make computations more efficient when necessary with algebraic multi-grid fast solvers, the Newton-Raphson method, an optimized GPU-accelerated FWI module, and in general various memory-saving techniques. \added{The coupled inversion technique---intrusive automatic differentiation (IAD)---as well as the software we developed, can be applied to other joint/coupled inversion problems as well. We hope the tools can stimulate further inverse modeling applications.}

\deleted[id=r0]{The applications of the intelligent automatic differentiation framework are by no means limited to the specific problem of fluid flow or \(\text{CO}_2\) storage. To facilitate further research, we release the codes for this specific project}
\begin{center}
    \deleted[id=r0]{\texttt{https://github.com/lidongzh/FwiFlow.jl}}
\end{center}
\deleted[id=r0]{and the library codes at }
\begin{center}
   \deleted[id=r0]{ \texttt{https://github.com/kailaix/ADCME.jl}}
\end{center}

Some work remains to be done for inverse modeling with coupled PDE systems and hidden dynamics of subsurface flow. First, the rock physics conversion requires accurate reference elastic properties, on which elastic properties are based and altered by fluid substitution. Inaccurate reference properties may lead to non-physical elastic properties at subsequent survey times, making it difficult to fit the physics-based PDEs. Thus, we strongly advise one to estimate the reference properties as accurately as possible before subsequent time-lapse inversion. Also, the so-called ``e factor" may for example depend on lithology. Second, seismic wave attenuation effects may be significant in flow-related problems. We can incorporate viscoelastic effects in wave physics modeling and expand elastic properties to anelastic properties, thus exploiting more wave physics behavior for flow inversion problems. Third, the gradient computation within the framework of automatic differentiation, which \texttt{ADCME} is based on, requires storing intermediate results for gradient back-propagation. In some situations, this will place excessive pressure on memory requirements. \replaced{Checkpointing}{Check-pointing} schemes can be used to alleviate this problem but require additional effort to implement. Finally, we must exercise care when interpreting the calibrated permeability or other hidden intrinsic parameters. If the flow has not reached certain areas throughout the slow time, the objective function has no sensitivity to flow parameters in those areas. \added[id=r0]{Due to the characteristics of the L-BFGS-B optimizer, there will be no update in places where the gradient is always zero. Therefore, one may want to examine the flow simulation with the inverted model to determine the areas valid for further analysis. One can also employ regularization methods as well as the multi-scale inversion strategy (i.e., to invert from low frequency to high frequency, from coarse parameterization to fine parameterization) to increase the robustness of inversion.}

These questions form a promising line of future research. While our extensive numerical investigation shows the applicability and effectiveness of this novel approach, answering these questions is critical for deeper understanding and more practical applications.
\begin{itemize}
    \item \added{First,} one such focused area we are pursuing is the investigation of the robustness of our coupled physics and \replaced{intrusive automatic differentiation}{intelligent AD} approach to sparse datasets, that is, sparsity in space and slow-time. This problem is essential when continuous subsurface monitoring strategies are needed, and full datasets are expensive to record at each slow-time step.
    \item  \added[id=r2]{Second, the inversion problem is more challenging in the presence of fractures. In this scenario, we need a reservoir simulator to handle the fractures, which requires a significant finer or an adaptive grid, and conduct automatic differentiation through this simulator. The resolution of hydrological properties that we reconstruct from seismic data in the presence of fractures also remains to be studied.} 
    \item \added[id=r0]{Third, the rock physics relationship can be heterogeneous in space. In our examples, we assume a homogeneous relationship as most coupled inversion practices, although we demonstrate that it is possible to calibrate the scalar parameter. In a future research direction, it is well worth considering inverting for a spatially variable rock physics parameter. This is again a multi-parameter inversion problem, where proper scaling may be critical, and alternate inversion of intrinsic parameters and rock physics parameters may also be necessary. The calibration of such a rock physics model with spatial variability can also be helpful for interpretation tasks.}
    \item  \added[id=r3]{Finally, we note that our approach, demonstrated here in 2-D, is fully expandable to more realistic settings in 3-D surface seismic, VSP seismic, and small-scale crosswell seismic surveys. Different geometries have different advantages and weaknesses. For example, the surface seismic surveys can monitor a large-scale reservoir, but the resolution is relatively low due to the limited high-frequency components in the data (up to tens of Hertz), and limited sampling of low-wavenumber components. On the contrary, small-scale crosswell surveys can generate data of frequency up to several thousand Hertz, sensitive to small-scale flow patterns, leading to a high-resolution delineation of the subsurface parameters, but it may not be able to monitor the whole reservoir. Practitioners can choose which geometry to use according to the problem at hand.}
\end{itemize}

\acknowledgments
The authors thank Biondo Biondi, Tapan Mukerji, and Gerald Mavko for many helpful discussions. Kailai Xu is supported by the Applied Mathematics Program within the Department of Energy (DOE) Office of Advanced Scientific Computing Research (ASCR), through the Collaboratory on Mathematics and Physics-Informed Learning Machines for Multiscale and Multiphysics Problems Research Center (DE-SC0019453). The source code of this project can be found at \url{https://github.com/lidongzh/FwiFlow.jl}, which generates all synthetic data used in this paper. We greatly appreciate the constructive and detailed comments and suggestions from associate editor Dr. Huisman and three anonymous reviewers.


%
%

\bibliography{GDP_FWI}

%
%
%
%
%

\end{document}